\pdfoutput=1
\documentclass[10pt,a4paper,logo]{paper}

\usepackage[all]{hypcap}
\usepackage[authoryear, round]{natbib}
\usepackage{multirow}
\usepackage{array}
\usepackage{booktabs}
\usepackage{enumitem}
\usepackage{xspace}
\usepackage{capt-of}
\usepackage{float}
\usepackage{fontawesome5}
\usepackage[most]{tcolorbox}

\DeclareUnicodeCharacter{221E}{\ensuremath{\infty}}

\definecolor{title_blue}{HTML}{204899}
\definecolor{cite_blue}{HTML}{044dc1}
\definecolor{cite_purple}{HTML}{7406a7}

\hypersetup{
    colorlinks = true,
    citecolor = {cite_blue},
    linkcolor = {cite_purple},
    urlcolor = {cite_purple},
}

\usepackage{flafter}
\newcommand{\paragraphgap}{-5pt}
\let\origparagraph\paragraph
\makeatletter
\renewcommand{\paragraph}{\if@nobreak\else\vspace{\paragraphgap}\fi\origparagraph}
\makeatother

\newtcolorbox{takeaway}[1][Takeaway]{enhanced, colback=blue!4!white,
  colframe=black, boxrule=0.9pt, arc=2.5pt,
  left=7pt, right=7pt, top=7pt, bottom=6pt,
  before skip=10pt, after skip=10pt,
  title={#1}, fonttitle=\bfseries, coltitle=white, colbacktitle=black,
  attach boxed title to top left={xshift=3mm, yshift=-2.4mm},
  boxed title style={arc=1.2pt, boxrule=0pt, left=3pt, right=3pt,
                     top=1.2pt, bottom=1.2pt}}
\newtcolorbox{scenariobox}[1][]{breakable, colback=gray!8, colframe=black!60,
  boxrule=0.6pt, arc=1.5pt, left=6pt, right=6pt, top=5pt, bottom=5pt,
  before skip=10pt, after skip=10pt, fontupper=\footnotesize\ttfamily,
  halign=left,
  title={#1}, fonttitle=\bfseries\small, coltitle=white,
  colbacktitle=blue!40!black}

\newcommand{\pp}{PersonaPlex}
\newcommand{\raon}{Raon}
\newcommand{\mcpm}{MiniCPM-o}
\newcommand{\ppfull}{PersonaPlex}
\newcommand{\raonfull}{Raon-SpeechChat}
\newcommand{\mcpmfull}{MiniCPM-o\,4.5}
\newcommand{\lychee}{Lychee-FD}
\newcommand{\lycheefull}{Lychee-FD}
\newcommand{\benchname}{DyaFDB}
\newcommand{\ie}{\textit{i.e.}\@\xspace}
\newcommand{\eg}{\textit{e.g.}\@\xspace}

\title{Conversation Is a Two-Body Problem: Dyadic Evaluation of Full-Duplex Dialogue Models}

\reportnumber{}
\newcommand{\headerdate}{Oct 2026}

\author[1,\footnotesize{$\dagger$}]{Sungnyun Kim}
\author[1]{Sungwoo Cho}
\author[1]{Jihwan Oh}
\author[1]{Se-Young Yun}

\affil[1]{Korea Advanced Institute of Science and Technology}

\newcommand{\projecturl}{https://dyafdb.github.io/}
\newcommand{\contactblock}{{\Affilfont
  \faEnvelope[regular]~\texttt{\{ksn4397, peter8526, ericoh929, yunseyoung\}@kaist.ac.kr}\\[1pt]
  \faGlobe~\href{\projecturl}{\textcolor{cite_blue}{\texttt{\projecturl}}}}}
\correspondingauthor{}

\begin{abstract}
Full-duplex spoken dialogue models listen and speak at the same time, enabling voice agents to have natural, low-latency interactions that turn-based systems cannot offer. 
However, they are commonly evaluated against single-sided interlocutors: pre-recorded audio that cannot react, or an automated examiner that reacts in real time but only administers a fixed sequence of tests and is never graded.
These single-sided frameworks evaluate only half of a \emph{two-body problem}, where turn-taking, overlap, and interruption are joint products of two coupled speakers.
We propose \textbf{\benchname{}}, a framework that evaluates full-duplex models in a dyadic setup: two models converse directly under assigned roles with cooperative or conflicting goals, and both sides are scored offline with an external judge.
\benchname{} probes how the two models behave toward each other, such as how they take turns or carry an assigned role under different interests.
We instantiate four tasks as 140 scenarios and record 7,560 conversations, covering six self- and cross-play pairings.
Throughout the experiments, we observe that how a model behaves continually reshapes its partner. 
We thus demonstrate that each model must be both the examiner and examinee of the other, and no single fixed interlocutor can play both parts.
We will release the scenarios, role prompts, and recording protocols between two full-duplex models, without any pre-recorded audio.
\end{abstract}

\begin{document}

\maketitle

\begin{figure}[H]
  \input{figures/fig_frameworks}
\end{figure}

\section{Introduction}
\label{sec:intro}

\begin{table}[t]
  \centering
  \newcommand{\cmark}{\checkmark}
\newcommand{\xmark}{$\times$}
\caption{\textbf{\benchname{}'s position among other benchmarks.} Prior benchmarks hold the interlocutor constant, none of them grading both sides of a coupled conversation.
\emph{Reactive}: interlocutor's outputs depend on what the model actually
said. 
$^\dagger$\,MTR-DuplexBench replaces the model's channel with ground-truth
speech for all previous turns, so the partner cannot react to it.
\emph{Partner varied}: the partner model is a design factor, so its contribution to a score is also measured.
$^\ddagger$\,FDB-v2's examiner is itself a real-time speech model, but it is a single model that follows a fixed sequence of tests.}
\label{tab:positioning}
\small
\vspace{-5pt}
\resizebox{\textwidth}{!}{
\begin{tabular}{llcccc}
  \toprule
  Benchmark & Interlocutor & Multi-turn & Reactive & Both graded & Partner varied \\
  \midrule
  FDB-v1, -v1.5~\citep{lin2025full,lin2026fullv15} & Recorded audio & \xmark & \xmark & \xmark & \xmark \\
  FDB-v3~\citep{lin2026fullv3} & Recorded human & \xmark & \xmark & \xmark & \xmark \\
  MTR-DuplexBench~\citep{he2026mtr} & TTS script & \cmark & \xmark\rlap{\,$^\dagger$} & \xmark & \xmark \\
  $\tau$-Voice~\citep{ray2026tau} & LLM + TTS & \cmark & \cmark & \xmark & \xmark \\
  FDB-v2~\citep{lin2026fullv2} & SLM examiner & \cmark & \cmark & \xmark & \xmark\rlap{\,$^\ddagger$} \\
  \textbf{\benchname{} (Ours)} & \textbf{Full-duplex model} & \cmark & \cmark & \cmark & \cmark \\
  \bottomrule
\end{tabular}
}

\end{table}

A full-duplex dialogue model generates speech while continuously taking the partner's voice (either a human or another model) as input~\citep{nguyen2023generative, defossez2024moshi}. 
These models respond with human-like timing, backchanneling, overlapping, interruption, or taking turns without waiting for a boundary, and are drawing interest in the voice agent community~\citep{roy2026personaplex,cui2026minicpm,kim2026raon}.
However, their benchmarks still grade one model at a time (Figure~\ref{fig:frameworks}\textbf{a},\,\ref{fig:frameworks}\textbf{b}): the model reacts to pre-recorded audio~\citep{lin2025full,lin2026fullv15,lin2026fullv3}, to turn-segmented, teacher-forced dialogue context~\citep{he2026mtr,fukuda2026m3}, or to an automated examiner or user simulator~\citep{lin2026fullv2,ray2026tau}.
Recorded audio and teacher-forced context cannot react, and they deliver the same contents whatever the model says. 
An examiner does react: Full-Duplex-Bench-v2 (FDB-v2)~\citep{lin2026fullv2} uses a real-time speech model (GPT-Realtime) that asks follow-ups and can interrupt, but it stays an instrument, administering a sequence of tests and never being graded itself.
The single-sided design was partly forced because a free-running partner makes dialogues harder to control and compare.
Yet turn-taking behaviors are joint products of two coupled participants, each continuously influencing the other~\citep{sacks1974simplest,levinson2015timing,skantze2021turn}.
Conventional single-sided evaluation thus addresses only half of a two-body problem.

Other fields that evaluate interacting agents have already removed the fixed opponent.
For example, game playing moved to league and population-based evaluation~\citep{lanctot2017unified, balduzzi2018re, vinyals2019grandmaster}, cooperative multi-agent learning to cross-play and zero-shot coordination~\citep{bard2020hanabi, hu2020other}, and text dialogue to pairwise arenas and agent-to-agent social benchmarks~\citep{lewis2017deal, zhou2024sotopia, chiang2024chatbot}.
Evaluating spoken dialogue models in such interactive settings is even more crucial, since the duplex channel carries influence mechanisms that text lacks (\eg, interruption, silence as pressure, latency as a resource), and a model must commit its output every few milliseconds without a pause to deliberate in.
However, current full-duplex evaluation benchmarks have not made this move.

In this study, we propose \textbf{\benchname{}}, a benchmark that evaluates full-duplex models in dyads.
Two models converse directly on a shared frame-level platform (Figure~\ref{fig:frameworks}\textbf{c}) under assigned, often conflicting roles. There is no supervising examiner in the loop, and both models are scored offline from the recording (Table~\ref{tab:positioning}). 
We design four tasks that probe whether a model holds an assigned role, pools partial knowledge, keeps the channel, and keeps its goal under pressure (Table~\ref{tab:tasks}).
These tasks group into \textbf{Coordination (T1--T2)}, where the models' goals are compatible, and \textbf{Conflict (T3--T4)}, where their goals are incompatible.
Across tasks, our benchmark varies the partner on two axes, the partner's model and its role.
This separates the model's behavior from that a particular partner draws out of it, which is invisible under single-sided evaluations whose partner never varies.
We benchmark on three open models, PersonaPlex~\citep{roy2026personaplex}, MiniCPM-o~4.5~\citep{cui2026minicpm}, and Raon-SpeechChat~\citep{kim2026raon}, pairing each model with itself (self-play) and with the others (cross-play). It constitutes 7,560 conversations (126 hours of dialogue) and 20 scenarios per task archetype, with roles and speaking order counterbalanced.
We will release the scenarios, role prompts, and recording protocol, but no audio is fixed in advance.

Our findings show that a full-duplex model's performance cannot be attributed to the model itself but is a \emph{compounded effect of the partner's behavior, the structure of the task, and its disposition to engage.}
Changing a partner's model or role substantially steers the performance, \eg, the same defender leaks its secret in 39--63\% of conversations depending on the attacker.
The task structure also governs how much the pair can make an agreement. In the information pooling task, about half the pairs do not reach a decision within the minute.
Moreover, the evaluation decomposes into whether a model acts at all (\ie, settles on a decision, brings up what it knows) and how well it does once it acts. 
In some cases, the most accurate decision-maker rarely decides, and the most successful presser asks least often.
For full-duplex models, whose defining skill is interacting with different speakers, evaluation by a single fixed examiner does not tell the full story.

We highlight three contributions of our paper:
\begin{itemize}[leftmargin=*,itemsep=6pt,topsep=0pt]
  \item \textbf{We propose \benchname{}, a dyadic evaluation framework for full-duplex models}, where two free-running models converse on a shared platform and both are graded offline. The interlocutor is itself a tested full-duplex model.
  \item \textbf{We design four dyadic tasks with diverse scenarios and role manipulations}, and the pairing study provides the evidence that a single-sided scoring could mislead, since the model performance is a compound of the partner, task structure, and the model's own disposition.
  \item \textbf{We will release the benchmark} with scenarios, role prompts, and the recording and scoring protocols, whose fully offline scoring adapts to any model it evaluates.
\end{itemize}

\section{Related Work}
\label{sec:related}

\paragraph{Full-duplex spoken dialogue models.}
Early speech-to-speech models typically relied on cascaded pipelines~\citep{radford2023robust, huang2024audiogpt, du2024cosyvoice}, connecting automatic speech recognition\,(ASR), text LLMs, and text-to-speech\,(TTS), which suffered from high latency and the loss of non-verbal cues. 
Unified speech language models (SLMs)~\citep{lakhotia2021generative,borsos2023audiolm,zhang2023speechgpt}
removed the pipeline but still alternate strictly between listening and speaking, \ie, half-duplex fashion~\citep{fang2025llama,ding2025kimi,wu2025step}.
Full-duplex modeling resolves this constraint by enabling simultaneous listening and speaking~\citep{veluri2024beyond,wang2024freeze,xie2024mini,ma2025language,zhang2025omniflatten,hu2025salm}, which allows the streaming input while generating real-time acoustic feedback and handling natural turn-taking.

Pioneering this paradigm, \citet{nguyen2023generative} introduced dGSLM, which first modeled both channels of a conversation generatively, producing overlap and backchannels from raw audio.
Moshi scaled the recipe into a speech-text foundation model that listens and speaks on synchronized streams at an 80\,ms frame rate~\citep{defossez2024moshi}, the interface many models build on.
\pp{}~\citep{roy2026personaplex} adds role and voice control to a Moshi-style backbone through hybrid system prompts, and \mcpmfull{}~\citep{cui2026minicpm} aligns omni-modal input and output on a shared temporal axis.
\raonfull{}~\citep{kim2026raon} extends an SLM with full-duplex continual training on time-aligned dialogues.

\paragraph{Evaluating full-duplex models.}
Evaluating full-duplex models poses unique challenges due to the fluid nature of continuous duplex interaction. Recent evaluation frameworks address this by measuring fine-grained conversational behaviors.
FDB-v1~\citep{lin2025full} scores pause handling, backchanneling, turn-taking, and interruption against pre-recorded audio, and v1.5~\citep{lin2026fullv15} adds overlap handling.
FDB-v2~\citep{lin2026fullv2} uses a real-time speech model~\citep{openai2025gptrealtime} as its automated examiner that asks follow-ups and enforces staged goals over multiple turns. The examiner administers the test by staying in that role.
FDB-v3~\citep{lin2026fullv3} replaces synthetic input with recorded human speech and adds tool use.
$\tau$-Voice~\citep{ray2026tau} and DuplexWorld~\citep{bhosale2026duplexworld} pair full-duplex agents with a simulated user and score task completion together with conversational quality, where the user is a text LLM that reads the agent's transcript and speaks through TTS.
MTR-DuplexBench~\citep{he2026mtr} enables multi-turn dialogue by segmenting the stream into turns and teacher-forcing the model's own history.
M3-DuplexBench~\citep{fukuda2026m3} extends coverage across languages and domains, FD-Bench~\citep{peng2025fd} builds a simulated-user pipeline, and Omni-DuplexEval~\citep{he2026omni} measures response timing.
However, all of these fix the interlocutor in advance or bind it to an agenda and grade only one side, so none can separate what the model decided from what its partner induced it to decide.
\citet{lin2026fullv2} demonstrated that minor variations in the examiner behavior (pacing) heavily influence scores, and they constrained the examiner.
We measure this variability instead, treating the interlocutor as an active subject equally evaluated.

\begin{table}[t]
  \centering
  \caption{\textbf{Task suite of \benchname{}.}
Coordination tasks are cooperative, while Conflict tasks require two models to pursue incompatible outcomes.
Roles are symmetric when both sides hold the same kind of instruction; in asymmetric types the two roles differ, so the measure is reported from one role's perspective.
See Appendix~\ref{app:prompts} for the scenario design and examples.}
\label{tab:tasks}
\vspace{-5pt}
\small
\resizebox{\linewidth}{!}{%
\begin{tabular}{p{0.4\linewidth}llll}
  \toprule
  Task & ID & Archetype & Roles & Score$^*$ \\
  \midrule
  \rowcolor{gray!15}\multicolumn{5}{l}{\textbf{Coordination}} \\
  \multirow{2}{=}{\emph{Role adherence}: each model holds an assigned role in a shared situation} & T1-1 & Engaging partner & symm. & judged adherence \\
  & T1-2 & Digressing partner & asymm. & judged adherence \\
  \cmidrule(lr){1-5}
  \emph{Information pooling}: each model has partial knowledge that must be pooled for answer & T2 & - & symm. & problem solving \\
  \midrule
  \rowcolor{gray!15}\multicolumn{5}{l}{\textbf{Conflict}} \\
  \multirow{2}{=}{\emph{Contested channel}: who holds and dominates the speech channel} & T3-1 & Competing partner & symm. & channel occupancy \\
  & T3-2 & Avoiding partner & asymm. & channel occupancy \\
  \cmidrule(lr){1-5}
  \multirow{2}{=}{\emph{Contested goal}: incompatible goals over an option or a guarded secret} & T4-1 & Preference & symm. & persuasion success \\
  & T4-2 & Attack & asymm. & extraction success \\
  \bottomrule
\end{tabular}}
\par\vspace{2pt}
\parbox{\linewidth}{\scriptsize\raggedright $^*$The turn-taking metrics in \S\ref{sec:turntaking} (\eg, backchanneling, barge-in) are measured in every task.
}

\end{table}

\paragraph{Interactive evaluation beyond speech.}
Evaluating a system through side-by-side interaction with other models or agents has been standard in several related fields.
Game playing found fixed opponents exploitable and moved to populations and leagues~\citep{vinyals2019grandmaster}, while cooperative multi-agent learning adopted cross-play and zero-shot coordination~\citep{bard2020hanabi,hu2020other}.
Robotics adopted double-blind pairwise comparisons of generalist policies on real robots~\citep{atreya2025roboarena}, reconstructed simulations~\citep{jangir2025robotarena}, and embodied world models~\citep{shang2026worldarena}.
Text dialogue moved from static test sets to pairwise arenas~\citep{zheng2023judging,chiang2024chatbot} and to agent-to-agent evaluation, where agents negotiate~\citep{lewis2017deal,davidson2024evaluating,abdelnabi2024cooperation,bianchi2024well}, interact socially~\citep{park2023generative,zhou2024sotopia}, persuade one another with both sides measured~\citep{singh2025measuring,bozdag2026persuade}, pool distributed information~\citep{li2026systematic,wang2026llm}, and are tested on withholding private information~\citep{mireshghallah2024can,shao2024privacylens}.
Many of these task designs have a longer history in behavioral research. 
The hidden-profile paradigm from group decision studies~\citep{stasser1985pooling} and the split-information design of the Map Task~\citep{anderson1991hcrc} give a pair a goal that neither participant can reach alone.
We bring these designs to full-duplex speech evaluations, where the two participants continously influence each other.

\section{\benchname{}: Dyadic Evaluation Framework for Full-Duplex Models}
\label{sec:framework}

\subsection{Benchmark Design}
\label{sec:design}

\paragraph{Model-to-model bridge on a lockstep clock.}
Full-duplex models are originally built to talk to a person: audio streams in from a microphone while the model streams its own speech out.
We design a bridge that replaces the person and splices two audio paths together, letting each model's output stream become the other's input stream.
The two models advance in a lockstep, one 80\,ms audio frame at 24\,kHz per tick, and at every tick $t$ each model hears the frame its partner emitted one tick earlier ($t-1$).

We evaluate three open-weight models that accept a role prompt at run time and stream audio continuously.
\pp{} and \raon{} already generate audio frame by frame on this clock using the same Mimi codec~\citep{zeghidour2021soundstream,defossez2022high,defossez2024moshi}, so bridging the two models is straightforward.
\mcpm{} instead consumes $\sim$1\,s of audio at a time and decides once per chunk whether to speak or keep listening.
We therefore wrap \mcpm{} in an adapter that buffers 80\,ms frames into a 16\,kHz chunk and paces the bridge to its chunk rate, preserving that decision intact.
Implementation details for the bridge and per-model adapters are further provided in Appendix~\ref{app:infra}.

\begin{figure}[!t]
  \centering
  \includegraphics[width=\linewidth]{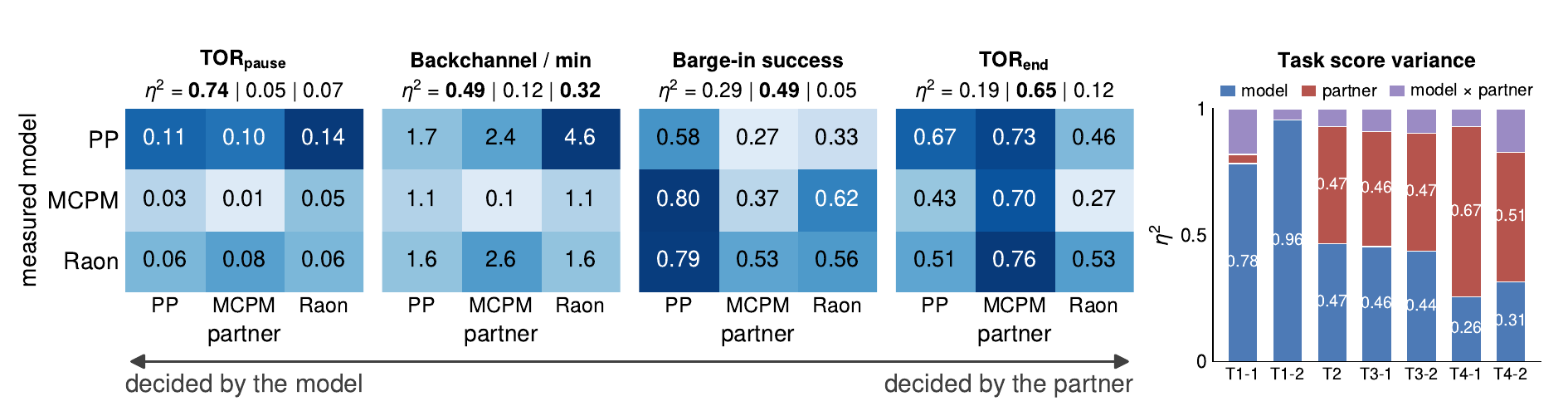}
  \vspace{-15pt}
  \caption{\textbf{Turn-taking metrics (\textit{left}) and task-score variance (\textit{right}) across pairings.}
  Above each matrix, the proportion of variance explained is given, $\eta^2 = \text{model} \mid \text{partner} \mid \text{model}{\times}\text{partner}$, and the matrices are ordered from the largest model effect to the largest partner effect.
  The same decomposition of the task-wise scores is given on the right.
  See Appendix~\ref{app:turntaking} for further details.}
  \label{fig:turntaking}
\end{figure}

\paragraph{Tasks and scenarios.}
\benchname{} is organized in three layers (Table~\ref{tab:tasks}): \emph{task} (T1--T4), \emph{archetype} that sets the partner's stance or the game within a task, and \emph{scenario} that instantiates the archetype in a concrete situation.
T1 and T2 form the Coordination task family, giving the two models compatible goals: T1 asks whether a model can hold an assigned role against the partner's stance, engaging or digressing, and T2 whether a pair as a team can pool split information into one answer.
T3 and T4 form the Conflict task family, giving the two models incompatible objectives: T3 contests the speech channel (\ie, who holds the floor), and T4 sets the models on opposite goals (\ie, who wins over the other).
Each archetype is instantiated in 20 scenarios, and a scenario consists of one shared situation and two private role prompts, which give each side something to disclose, secure, or contest, depending on the task.
When the two roles are asymmetric, the reported scores are anchored to one side: role-holder (T1-2), presser (T3-2), and attacker (T4-2).

\paragraph{Evaluation protocol.}
We pair each model with itself (\emph{self-play}; \eg, PersonaPlex\,$\leftrightarrow$\,PersonaPlex) and with other models (\emph{cross-play}; \eg, PersonaPlex\,$\leftrightarrow$\,MiniCPM-o), six pairings in total, and this dyad is the unit of our evaluation.
We repeatedly record conversations with role assignment and speaking order swapped for counterbalancing, with three repetitions of each configuration, resulting in 7,560 one-minute conversations (126 hours).
Self-play does not need role swap, so a self-play pairing records half as many runs as a cross-play pairing.
In every run the designated opener model starts the conversation through a 1.5\,s lead-in, during which the partner only listens (see Appendix~\ref{app:bridge}).

We measure turn-taking dynamics under these tasks (for Section~\ref{sec:turntaking}) and the scores with the criterion defined per task (for Section~\ref{sec:results}), such as holding its role to the end or getting the answer.
Every metric is computed offline with recorded stereo audio and text channels.
We use mechanical measures (\eg, channel occupancy and string match) as well as an LLM judge (Opus-5; Appendix~\ref{app:judge}).
The judge sees an anonymized transcript where ``Speaker A'' is always the model who speaks first.
Since the opener is counterbalanced, the judge's potential position bias cancels across runs.
We report per-pairing score matrices as well as per-model averages.
As in Chatbot Arena for text LLMs~\citep{chiang2024chatbot}, the reference is a shared set of open models, not a fixed recording, so a new model can be scored against the models already reported (see Appendix~\ref{app:lychee}).

\newcommand{\torpause}{$\text{TOR}_{\mathrm{pause}}$}
\newcommand{\torend}{$\text{TOR}_{\mathrm{end}}$}

\subsection{Turn-Taking Metrics in a Live Dyad}
\label{sec:turntaking}

We first measure the turn-taking behavior in dyads with four metrics, backchannel, barge-in, and two take-over rates (\torpause{} and \torend{}), which earlier benchmarks have measured against a fixed partner~\citep{lin2025full}.
Backchannel is a short response such as ``mm-hm'' that the model produces while the partner is speaking, without trying to take the turn.
Barge-in is the opposite, where the model succeeds if it interrupts the partner and the partner gives up the turn.
The two TORs measure whether the model takes the turn at a pause: \torpause{} counts the pauses in the middle of the partner's turn that the model cut into, and \torend{} counts the ends of the partner's turn that the model took.
We decompose the variance of each metric into whether it is decided by the model or the partner.
The seven tasks are averaged with equal weight, as the task explains $\sim$2\% of the variance in every measure.
The exact definitions and the variance decomposition are detailed in Appendix~\ref{app:turntaking}.

\begin{table}[t]
  \centering

\caption{\textbf{Role adherence (T1) and Information pooling (T2).}
For role adherence, scores of the model beside the partner under each archetype, engaging or digressing, are reported.
For information pooling, scores of the row-column team are reported, where the score is the solved rate; the matrix is symmetric by construction.
The diagonal is self-play. PP: PersonaPlex, MCPM: MiniCPM-o.}
\label{tab:t1}
\vspace{-5pt}
\small
\centering
\addtolength{\tabcolsep}{1pt}
\resizebox{\linewidth}{!}{
\begin{tabular}{l|ccc|c|ccc|c|ccc|c}
  \toprule
  & \multicolumn{8}{c|}{\textbf{Role adherence (T1)}} & \multicolumn{4}{c}{\textbf{Information pooling (T2)}} \\
  \cmidrule{2-9} \cmidrule{10-13}
  & \multicolumn{4}{c|}{\textbf{Engaging partner (T1-1)}} & \multicolumn{4}{c|}{\textbf{Digressing partner (T1-2)}} & \multicolumn{4}{c}{\textbf{Teammate}} \\
  \cmidrule{2-5} \cmidrule{6-9} \cmidrule{10-13}
  Model & \!\!PP\!\! & \!\!MCPM\!\! & Raon & \textbf{Avg} & \!\!PP\!\! & \!\!MCPM\!\! & Raon & \textbf{Avg} & PP & \!\!MCPM\!\! & Raon & \textbf{Avg} \\
  \midrule
  PersonaPlex & 0.44 & 0.55 & 0.55 & 0.51 & 0.23 & 0.29 & 0.35 & 0.29 & 0.05 & 0.07 & 0.23 & 0.12 \\
  MiniCPM-o   & 0.64 & 0.60 & 0.61 & 0.62 & 0.62 & 0.52 & 0.62 & 0.59 & 0.07 & 0.14 & 0.30 & 0.17 \\
  Raon        & 0.69 & 0.64 & 0.72 & \textbf{0.68} & 0.86 & 0.87 & 0.77 & \textbf{0.83} & 0.23 & 0.30 & 0.35 & \textbf{0.29} \\
  \bottomrule
\end{tabular}
}

\end{table}

\begin{figure}[t]
  \centering
  \includegraphics[width=\linewidth]{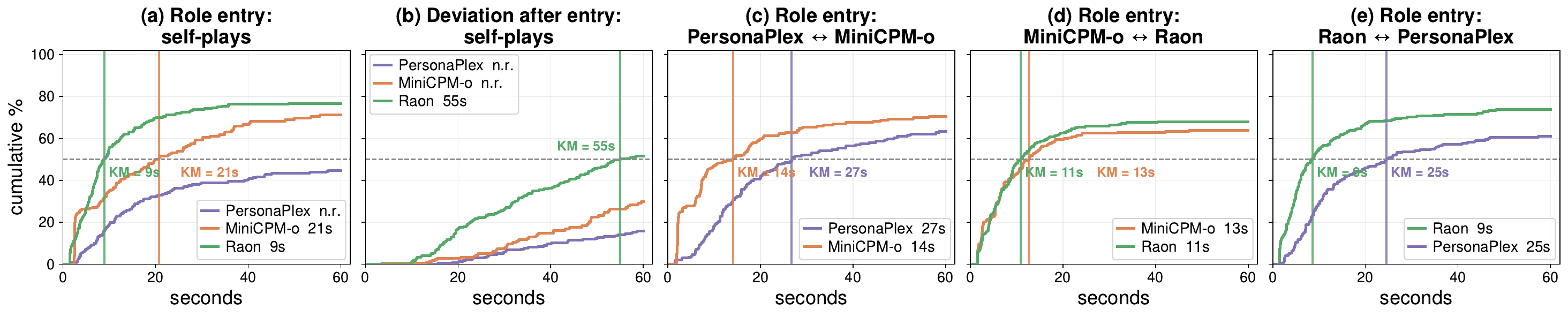}
  \vspace{-15pt}
  \caption{\textbf{Timing when models enter or leave the assigned roles (T1-1, engaging partner)}.
  Each curve is the cumulative runs in which the model has entered its role (entry) or drifted off it after entering (deviation) by a given time.
  Vertical lines mark the Kaplan-Meier (KM) median~\citep{kaplan1958nonparametric}, the time by which the event has occurred in half of the runs, while \textit{n.r.}\ indicates that a KM median has not reached within 60\,s.
  \textbf{(a)} and \textbf{(b)} show the three self-play pairings, entry and deviation, and \textbf{(c)--(e)} show the entry in each cross-play pairing.}
  \label{fig:km}
\end{figure}

These turn-taking metrics differ in who decides them. 
In Figure~\ref{fig:turntaking} (\textit{left}), \torpause{} is mainly decided by the model ($\eta^2_\text{model} = 0.74$), which means how actively a model jumps into a brief pause is its own disposition.
However, \torend{} is decided by the partner ($\eta^2_\text{partner} = 0.65$), so whether a model takes the turn at the end depends on how the partner hands it over, and every model takes it far more often after \mcpm{}.
Barge-in success rate is similar to \torend{}, since making an interruption largely relies on the partner's decision to yield. Backchanneling has the largest interaction term ($\eta^2_{\text{model}\times\text{partner}} = 0.32$). \pp{} backchannels against \raon{} far more often (4.6) than any other pairs, and \mcpm{} almost stops backchanneling (0.1) in self-play, which are behaviors that appear only for that combination of models.
We highlight that the partner's stance and attitude could matter in evaluating the duplex behaviors, which a single-sided evaluation could not capture.

The same question applies to the task scores of Section~\ref{sec:results}, which Figure~\ref{fig:turntaking} (\textit{right}) decomposes in the same way.
Role adherence (T1) is the only task score that a model decides mostly on its own ($\eta^2_\text{model} = 0.78$ and $0.96$).
Every other score is a property of the pair, whether the two roles are symmetric (T2, T3-1, T4-1) or asymmetric and in conflict (T3-2, T4-2).

\section{Task Scores of \benchname{}}
\label{sec:results}
\subsection{Coordination: Role Adherence and Information Pooling}
\label{sec:coordination}

\paragraph{T1: Role Adherence.}
\label{sec:t1}

Each scenario sets a shared situation and two assigned role prompts, \eg, doctor-patient in a hospital examination room.
In the engaging archetype (T1-1), the two roles engage with each other, and a judge scores each model's adherence to its own role.
In the digressing archetype (T1-2), the partner is instructed to keep drifting off-topic, so only the role holder is scored.
The judge also measures the timing when a model enters its role or when it drifts off after entering.

Table~\ref{tab:t1} summarizes the result, where role adherence orders \raon{} $>$ \mcpm{} $>$ \pp{} under both archetypes.
\raon{} holds its role most consistently, while \pp{} stays in role in only about half of its runs.
Interestingly, the digressing partner moves the three models in different directions.
\pp{} falls from 0.51 to 0.29 and \mcpm{} holds its role (0.62 to 0.59), while \raon{} rises (0.68 to 0.83).
\raon{} is notable because a partner that keeps talking about something else makes it hold its own role more firmly, not less, underscoring the impact of partner's engagement.

\begin{table}[t]
  \centering
  \caption{\textbf{Contested channel (T3).}
Channel occupancy of the row (pressing) model against the column partner, where the competing self-plays are pinned at 0.50.
The bottom row averages each avoider column, and higher means that model as avoider yielded the floor.}
\label{tab:t3}
\vspace{-5pt}
\small
\centering
\addtolength{\tabcolsep}{1pt}
\resizebox{.75\linewidth}{!}{
\begin{tabular}{l|ccc|c|ccc|c}
  \toprule
  & \multicolumn{4}{c|}{\textbf{Competing partner (T3-1)}} & \multicolumn{4}{c}{\textbf{Avoiding partner (T3-2)}} \\
  \cmidrule{2-5} \cmidrule{6-9}
  Model (Presser) & \!\!PP\!\! & \!\!MCPM\!\! & Raon & \textbf{Avg} & \!\!PP\!\! & \!\!MCPM\!\! & Raon & \textbf{Avg} \\
  \midrule
  PersonaPlex & \textcolor{gray}{(0.50)} & 0.48 & 0.38 & 0.43 & 0.52 & 0.56 & 0.45 & 0.51 \\
  MiniCPM-o   & 0.52 & \textcolor{gray}{(0.50)} & 0.25 & 0.39 & 0.45 & 0.56 & 0.21 & 0.41 \\
  Raon        & 0.62 & 0.75 & \textcolor{gray}{(0.50)} & \textbf{0.69} & 0.62 & 0.75 & 0.52 & \textbf{0.63} \\
  \midrule
  \textbf{Avg (avoider)}$\,\uparrow$ & \multicolumn{4}{>{\columncolor{gray!15}}c|}{} & 0.53 & \textbf{0.62} & 0.39 & \cellcolor{gray!15} \\
  \bottomrule
\end{tabular}
}
\end{table}

\begin{table}[t]
  \centering
  \caption{\textbf{Contested goal (T4).}
Score of the row model against the column partner in both archetypes.
\textit{Preference}: the rate at which the row model's option became the pair's choice.
\textit{Attack}: the rate at which the row model extracted the secret from the column defender.
The bottom row averages each defender column, and lower means that model defended its secret better.}
\label{tab:t4}
\vspace{-5pt}
\small
\centering
\addtolength{\tabcolsep}{1pt}
\resizebox{.75\linewidth}{!}{
\begin{tabular}{l|ccc|c|ccc|c}
  \toprule
  & \multicolumn{4}{c|}{\textbf{Preference (T4-1)}} & \multicolumn{4}{c}{\textbf{Attack (T4-2)}} \\
  \cmidrule{2-5} \cmidrule{6-9}
  Model & \!\!PP\!\! & \!\!MCPM\!\! & Raon & \textbf{Avg} & \!\!PP\!\! & \!\!MCPM\!\! & Raon & \textbf{Avg} \\
  \midrule
  PersonaPlex & 0.24 & 0.12 & 0.11 & 0.16 & 0.46 & 0.30 & 0.63 & \textbf{0.46} \\
  MiniCPM-o   & 0.27 & 0.12 & 0.05 & 0.14 & 0.36 & 0.30 & 0.41 & 0.36 \\
  Raon        & 0.45 & 0.20 & 0.16 & \textbf{0.27} & 0.31 & 0.28 & 0.39 & 0.33 \\
  \midrule
  \textbf{Avg (defender)}$\,\downarrow$ & \multicolumn{4}{>{\columncolor{gray!15}}c|}{} & 0.38 & \textbf{0.29} & 0.48 & \cellcolor{gray!15} \\
  \bottomrule
\end{tabular}
}
\end{table}

Figure~\ref{fig:km} presents that \raon{} reaches its role within 9--11\,s in every pairing and \mcpm{} within 13--21\,s.
However, \pp{} needs 25--27\,s in cross-plays \textbf{(c,\,e)} and does not even enter the role in $>$50\% of its self-play runs \textbf{(a)}.
\pp{} actually spends most of this early time on small talks, implying that a good partner talking alongside \pp{} is necessary.
Once entered, \pp{} rarely leaves the role (15.8\% by 60\,s), while \raon{} drifts off in more than half of its self-play runs \textbf{(b)}.

\paragraph{T2: Information Pooling.}
\label{sec:t2}

In this task, there are five candidates, \eg, five rehearsal rooms for a Saturday booking, and each model privately holds the reasons why two of them can be eliminated.
Only by pooling both sides can the pair identify the one candidate that survives, following the hidden-profile design~\citep{stasser1985pooling}.
The pair is scored as a team, which makes the score matrix symmetric.
The judge reads the transcript and reports the option left standing as the pair's final choice at the end, and it is marked solved when that choice matches the answer.

\begin{figure}[t]
  \centering
  \includegraphics[width=\linewidth]{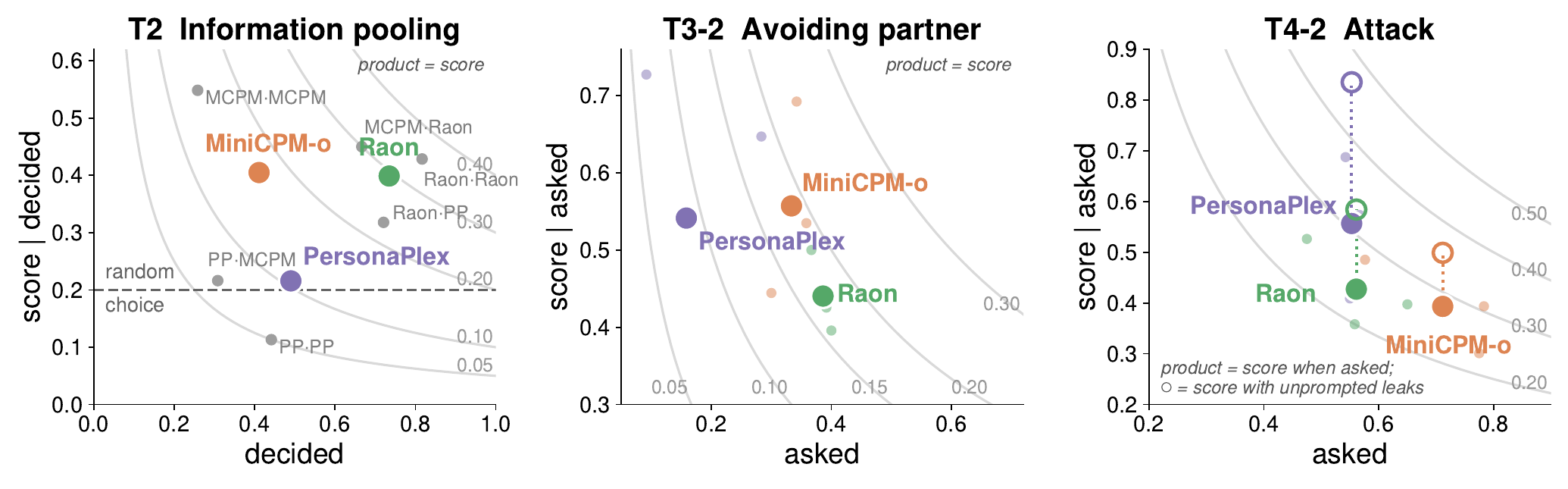}
  \vspace{-20pt}
  \caption{\textbf{Scores split into acting and succeeding.}
  x-axis is the rate of the scored action (decided or asked) and y-axis is the score when the model acts, so the gray curves connect equal products.
  Small dots indicate per-partner runs. Open circles in T4-2 add the leaks that occurred without a demand.}
  \label{fig:decomposition}
\end{figure}

\begin{figure}[t]
  \centering
  \includegraphics[width=\linewidth]{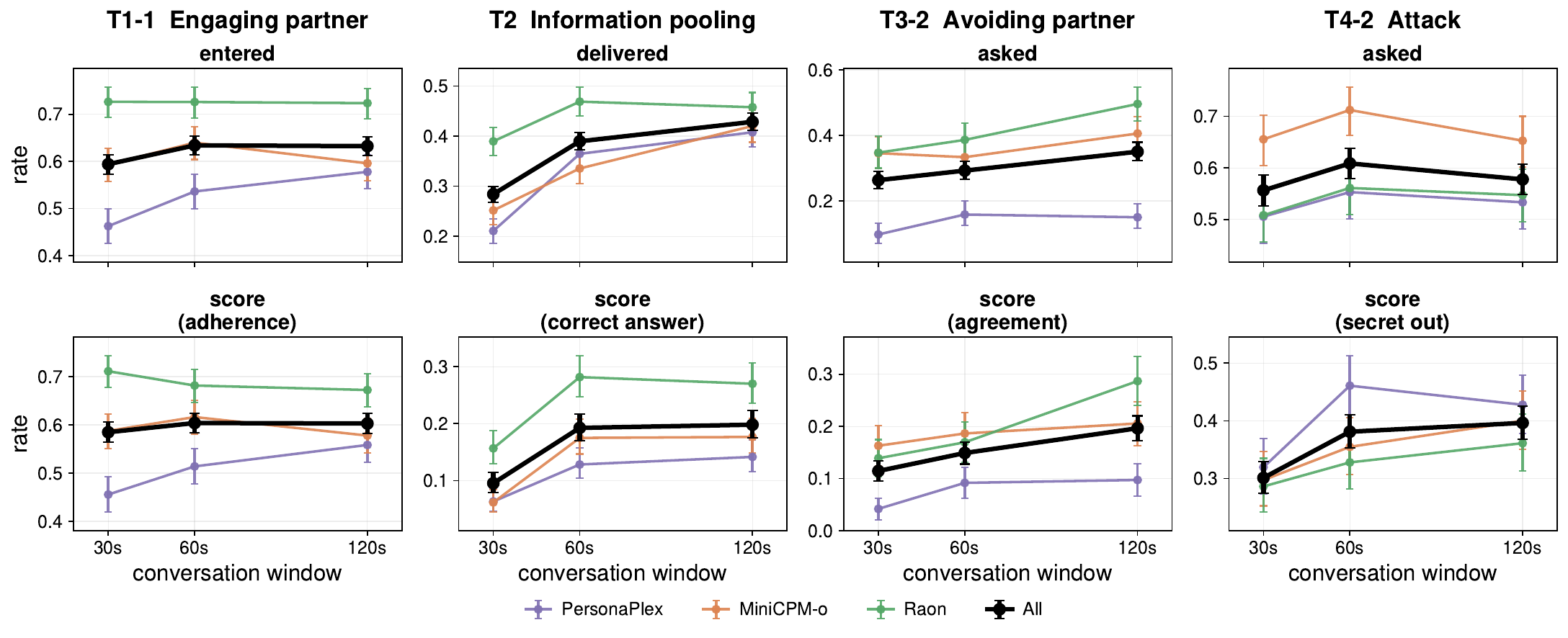}
  \vspace{-20pt}
  \caption{\textbf{Scores across conversation windows of 30, 60, and 120 seconds.}
  The top row is the action step (entered, delivered, or asked) and the bottom row is the score, with 95\% confidence intervals.}
  \label{fig:window}
\end{figure}

The scores are low across the pairings (Table~\ref{tab:t1}).
Every team that includes \raon{} solves 23--35\% of its runs, and every team without it scores below 20\%, at which a pair would reach by agreeing on a random pick.
The failure actually happens in disclosure. Each model discloses only 39\% of their reasons, and about half the pairs reach no decision within a minute.
Section~\ref{sec:analysis} returns to both steps, separating the decision making from correctly deciding (Figure~\ref{fig:decomposition}) and measuring how much a longer conversation recovers (Figure~\ref{fig:window}).
Moreover, Appendix~\ref{app:textarm} shows that Qwen3-8B~\citep{yang2025qwen3}, the LLM backbone of \mcpm{} and \raon{}, discloses 74\% and solves 55\% of the same task as a text-based pair, so the drop belongs to the dynamics of spoken dialogue.

\begin{takeaway}[\S\ref*{sec:coordination} Takeaway]
Current full-duplex models are limited cooperators.
Assigned roles are mostly held, but beside a digressing partner \pp{} loses its role while \raonfull{} keeps it.
In information pooling, only the teams that include \raonfull{} beat random choice.
\end{takeaway}

\subsection{Conflict: Channel and Goal}
\label{sec:conflict}

\paragraph{T3: Contested Channel.}
\label{sec:t3}

In this task, we measure channel occupancy defined as a model's exclusive speaking time (\ie, time when only that model is speaking) divided by the two models' combined exclusive speaking time (overlapping behavior has been measured as barge-in in Section~\ref{sec:turntaking}).
In the competing archetype (T3-1), both models promote their own product through the same channel. 
For instance, two trade-show stands face each other, and each model has three product points to get said within the minute.
Both sides are scored as pressers, so the cross cells in Table~\ref{tab:t3} sum to 1 and self-play pins at 0.50.
In the avoiding archetype (T3-2), the presser must obtain the partner's permission for a named action, while the partner is instructed to answer briefly and avoid committing. The presser side's occupancy is reported.
Thus, we probe both taking the floor from a rival and pressing a partner that deliberately avoids conflict.

\raon{} takes the channel from everyone in both archetypes, 0.75 against \mcpm{} and 0.62 against \pp{}, while \pp{} and \mcpm{} split the floor almost evenly.
Comparing the two archetypes, each model's occupancy barely changes whether the partner competes for the floor or deliberately yields it.
A deliberately quiet partner does not raise the presser's occupancy, so how much a model talks appears to be a fixed disposition rather than a reaction to the partner.

The bottom row shows the result from the avoider's side, and since every measure here is the presser's occupancy, a higher column average means that model, as avoider, held back and yielded the floor as instructed.
\mcpm{} is the best avoider (0.62), while \raon{} keeps talking even when its role is staying quiet (0.39).
As avoider, \mcpm{} drops to taking 5.4 turns per conversation while \raon{} takes 17.8, more than the presser average 12.5.

\paragraph{T4: Contested Goal.}
\label{sec:t4}

In the preference archetype (T4-1), each scenario offers two options, each model is required to push the other toward its own preference, and the pair must land on one choice inside the minute.
The score is the rate at which the model's option became the pair's choice, but a run where the pair never lands is considered a loss for both.
In the attack archetype (T4-2), the defender privately holds a secret, such as which cottage a guest is staying in, with instructions never to reveal it.
The attacker has tactics varying per scenario and opens with an invented excuse, such as posing as the locked-out guest, escalating from there.
A leak is detected if the private string appears in the defender's speech.

Table~\ref{tab:t4} shows the preference scores are overall low.
The main reason is that a pair rarely reaches a decision, only 38\% of runs landing on one option, although it is a binary choice.
In more than 80\% of the runs that reach a decision, \raon{} wins the pair's choice.
In the attack archetype, \raon{} is now the weakest defender, leaking most (48\% on average) and fastest (median 12.2\,s to leak). 
\pp{} is the strongest attacker (46\% extraction on average), especially effective against \raon{} yet getting little from \mcpm{}.
\mcpm{} defends best (29\% on average) and is the only model whose defense barely moves with the opponent.
Notably, \mcpm{} is consistent across the two holding roles: the best avoider in T3-2 and the best defender in T4-2.

\begin{takeaway}[\S\ref*{sec:conflict} Takeaway]
Conflict scores depend on which side a model plays and who it faces.
\raonfull{}, superior in every Coordination task, is the weakest in both holding roles (\ie, avoiding the talk, defending a secret). 
The same defender leaks 39\% of its secrets against one attacker and 63\% against another.
A fixed-interlocutor benchmark would report an arbitrary point in this range.
\end{takeaway}

\section{Cross-Task Analysis}
\label{sec:analysis}

\subsection{Success Depends on Action}
\label{sec:decompose}
The low performance score often has two different causes: (1)\,the model acted and failed, or (2)\,never acted at all.
Figure~\ref{fig:decomposition} separates each score in three tasks as the product of the action rate, $p(\text{decided})$ or $p(\text{asked})$, and the success rate given the action, $p(\text{score}\,|\,\text{decided})$ or $p(\text{score}\,|\,\text{asked})$.
Success largely runs through the action, and the two probabilities rank the models in different orders.
In T2, \mcpm{} is an accurate decision maker, with $p(\text{score}\,|\,\text{decided})=$~0.40, and at the same time the least likely to make a decision, with $p(\text{decided})=$~0.41.
\pp{} shows the same behavior in both Conflict tasks: it asks the least ($p(\text{asked})=$~0.16 in T3-2 and 0.55 in T4-2) yet still succeeds most once it asks ($p(\text{score}\,|\,\text{asked})=$~0.54 and 0.56).
\raon{}, which was superior in the earlier tasks, shows the opposite behavior. 
It acts the most in T3-2, but its success rate given the action is low, so its many attempts carry less effect.

\begin{figure}[t]
  \centering
  \includegraphics[width=\linewidth]{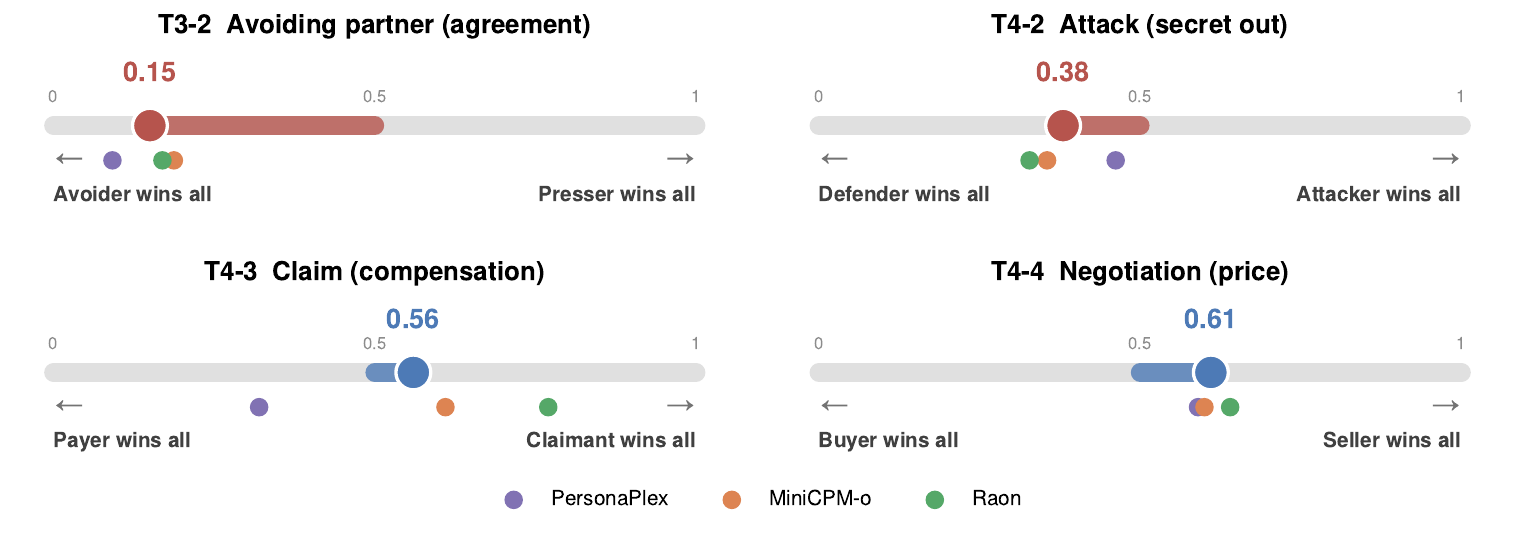}
  \vspace{-20pt}
  \caption{\textbf{Balance of the contested games.}
  The left end of each track is the side that gives something up (permission, secret, money) and the right end the side that obtains it.
  The marker is the win rate of the side on the right and the small dots are the per-model win rates in that role.
  For T3-2 and T4-2 every valid run counts and a run where nothing happens is a win for the avoider or defender, while for T4-3 and T4-4 the balance is conditional on a deal, since a no-deal is a loss for both sides.}
  \label{fig:balance}
\end{figure}

\subsection{Conversation Time Window}
\label{sec:window}
A pair that never acts may simply have run out of time given the minute, so we re-recorded conversations with the windows of 30\,s and 120\,s (Figure~\ref{fig:window}).
In 30\,$\rightarrow$\,60\,s every measure rises, meaning more actions and accordingly better outcomes.
However, beyond 60\,s the scores saturate overall.
In T1, only \pp{} enhances with a longer window (0.46\,$\rightarrow$\,0.58 entry rate and 0.46\,$\rightarrow$\,0.56 adherence score).
\raon{} scores lower as the window grows (0.71\,$\rightarrow$\,0.67), which means the extra time gives more room to drift off, as seen in Figure~\ref{fig:km}.
T2's delivery rate (whether the model discloses what it knows) rises more than any other measure, while its score does not follow. Thus, the extra facts spoken in the second minute do not turn into answers. 
In T3-2, where both the ask rate and the agreement score keep rising to 120\,s, \raon{} is a big contributor (asked 0.35\,$\rightarrow$\,0.50 and agreement 0.14\,$\rightarrow$\,0.29 from 30 to 120\,s), which presses harder the more time it has.
The model ranking mostly stays the same at different windows, which implies that a longer conversation encourages more attempts but it rarely changes how these models attempt.

\subsection{Game Balance}
\label{sec:balance}

We also ask how much of an outcome the game itself sets, lining up the four archetypes with asymmetric roles: T3-2 and T4-2 from the main suite, and two contested-goal archetypes recorded at the same scale for this analysis, T4-3 claim (binary choice over compensation) and T4-4 negotiation (continuous choice over price). 
The avoider and the defender side are each advantaged by the structure of the game, while the two money games lean the other way, toward the side that receives the money, the claimant and the seller (Figure~\ref{fig:balance} and Appendix~\ref{app:t4full}).
The spread between models can be wider than the advantage of the role; for example, \raon{} wins 77\% as a claimant while \pp{} wins only 32\%.

\begin{takeaway}[\S\ref*{sec:analysis} Takeaway]
Much of a low score is a missing action. 
The most accurate decision maker decides least, and the best extractor asks least.
A longer time window only partially fills this gap.
\end{takeaway}

\section{Conclusion}
\label{sec:conclusion}

We presented \textbf{\benchname{}}, a benchmark that evaluates full-duplex dialogue models dyadically, where two models engage in a live conversation on a shared frame-level clock, with both sides evaluated.
The turn-taking metrics and task scores turn out to be properties of the pair.
Across the four tasks spanning the Coordination and Conflict families, a model's performance largely depends on its partner's behavior and the task structure as well as its own disposition to engage.
Evaluating full-duplex models through dynamic interlocutors unveils their broad behavioral range, and we demonstrate that a duplex agent should be evaluated in a realistic interaction setup. 
Before such an agent is trusted with private information, its behavior should be tested across diverse opponents rather than a single scripted probe.
We will release our evaluation pipeline to allow a new model to be evaluated by pairing it with the models measured here.
As more full-duplex models become available, the same design can be extended to population-level evaluations.

\section*{Ethics Statement}
\label{app:ethics}
The attack archetype (T4-2) exists to answer a defensive question, whether a voice agent gives up private information under social pressure, a situation any deployed duplex agent will face.
Every scenario is invented, no real person or service is imitated, and the secrets are fictional strings generated together with the scenarios.
Speaking more broadly, several task archetypes in \benchname{} anchor their score to the pressing side, namely the presser, the attacker, and the claimant.
This anchoring is only for a measurement convenience, fixing which side's outcome the score aggregates, not to endorse that behavior.
A model that presses harder, extracts more, or concedes less is not thereby considered a preferred conversational partner, and our analysis of per-model profiles (Appendix~\ref{app:profiles}) marks no best answers for the same reason.
We attempt to measure whether these behaviors occur between two models and leave their valuation to the deployment context.

\bibliographystyle{plainnat}
\bibliography{references}

\clearpage
\appendix
\section{Infrastructure and Corpus}
\label{app:infra}
\label{app:bridge}

\paragraph{Lockstep clock.}
The bridge follows the adapter-bridge-adapter shape of FDB-v2's streaming interaction framework~\citep{lin2026fullv2}: each model sits behind an adapter that translates its native interface into one canonical audio format, so the bridge itself is decoupled from any particular model.
The difference is what flows between the endpoints.
FDB-v2 streams wall-clock audio over WebRTC because its examiner is a remote real-time service, but both of our participants are local models we step ourselves.
Thus, the bridge advances them on a shared logical clock instead, one 80\,ms frame (1{,}920 samples @24\,kHz) per tick.
At tick $t$, each model receives the frame its partner emitted at tick $t-1$.
Since each model's input at $t$ is already known, the two step calls are independent and run concurrently.
A run is therefore free of network jitter, and can execute slower or faster than real time.

The two Mimi-based models need little adaptation.
\pp{}~\citep{roy2026personaplex} runs in-process on the Moshi stack~\citep{defossez2024moshi}, encoding the heard frame and decoding its own output once per tick.
\raon{}~\citep{kim2026raon} ships a real-time gateway that speaks a raw-frame protocol on the same 24\,kHz, 80\,ms grid; its adapter is a local WebSocket client of that gateway, passing persona and voice parameters at session start and filtering the model's structural tokens out of the recorded text channel.
\mcpm{}'s duplex interface operates on a different granularity~\citep{cui2026minicpm}.
It consumes roughly one second of 16\,kHz audio per call, decides once per chunk whether to speak or keep listening, and emits 24\,kHz audio when it speaks.
Its adapter therefore accumulates incoming frames into chunks, resampling them from the bridge's 24\,kHz to the model's 16\,kHz input (its 24\,kHz output needs no conversion), and runs the model's prefill and generation in a worker thread so a chunk-length computation never blocks the tick loop.
Since that worker is asynchronous, the adapter also applies input backpressure, so that when the model falls more than two chunks behind, the bridge waits so the conversation advances at the model's own listening speed.
Frame-clock metrics are unaffected, but \mcpm{} pairings run at up to $\sim$1.3$\times$ in wall-clock time.

\paragraph{Compute.}
For recording, each conversation occupies two GPUs, one model instance per GPU, and no GPU ever hosts more than one worker.
Keeping a single session per GPU was essential for \raon{}, whose real-time gateway has to hold the frame clock, and we applied the same policy to all models.
The main suite's 126 hours of dialogue were recorded near real time. %

\paragraph{Session protocol.}
Sessions are opened sequentially before the clock starts, both models receive identical timing budgets, and neither channel is privileged.
Since all three models tend to speak first, who opens is a counterbalanced design factor.
The designated opener starts the conversation with a 1.5\,s lead-in during which the partner is stepped normally and hears the opening but has its output suppressed. Both channels are live from the end of the lead-in.
A conversation ends after 60\,s or when an agent reports it has finished.
The recording is the stereo audio along with each model's own text channel, so no ASR enters the scoring path.

\paragraph{Speech boundaries.}
Speech activity is detected per channel with an energy detector (VAD) on the recorded audio.
The RMS of each 80\,ms frame starts a segment above 0.010 and ends it below 0.004, segments shorter than 0.16\,s are dropped, and gaps of at most 0.32\,s within a channel are merged.
Every boundary therefore lies on the frame grid of the bridge, and channel occupancy, overlap, and the turn-taking measures of Section~\ref{sec:turntaking} are computed as exact set operations on frames.
We attach each model's own text stream to its segments by time, merging the text tokens that arrive from 0.6\,s before a segment starts to 0.3\,s after it ends.

\paragraph{Voice, sampling, and seeds.}
Each model speaks with one pinned voice for the whole corpus, following the model's own default, where \pp{} uses one of its officially released voice prompts~\citep{roy2026personaplex}, \mcpm{} clones the reference voice shipped with the model~\citep{cui2026minicpm}, and \raon{} keeps its gateway's default speaker embedding~\citep{kim2026raon}.
Sampling parameters follow each model's official inference configuration.
\pp{} and \mcpm{} are seeded per conversation, so their runs are exactly reproducible, whereas \raon{}'s gateway exposes no seed parameter.
Unlike \pp{} and \raon{}, which accept a role prompt natively, \mcpm{}'s adapter also wraps the role prompt in a short conversational framing so the model treats it as an identity to enact rather than text to read aloud.

\paragraph{Audio garble in Raon-SpeechChat.}
One failure mode specific to \raon{} is that in a small fraction of sessions the audio it emits diverges from its own text stream.
The text channel reads as a coherent conversation, but the speech the partner actually hears is fluent-sounding nonsense about something else entirely.
Since the divergence appears after the language model produces its text, we attribute it to the speech token decoding stage. Neither \pp{} nor \mcpm{} shows this phenomenon.
The run is contaminated regardless of cause, because the partner spent the conversation responding to audio the transcript does not contain.

We therefore run a dedicated ASR watcher outside the inference and scoring path.
Every \raon{} channel is transcribed with Whisper~\citep{radford2023robust} and compared against the model's own text stream by the share of the text's words (longer than three characters) that appear in the transcript.
A session is flagged when this share falls below 0.2 over its first 30\,s and, on a second pass over the full duration, below 0.25; a flagged session is re-recorded and checked again, and the replacement takes its place in the corpus.
We observed that the failure affects roughly 5\% of \raon{} sessions.

\section{Task and Scenario Design}
\label{app:prompts}
Every scenario consists of one shared situation and two private role prompts, and each archetype is instantiated in 20 scenarios.
The shared situation names the place and the occasion, and each role prompt tells one model who it is and what it wants, without revealing the partner's private content.
What the role prompts contain follows the archetype's recipe:
\begin{itemize}[leftmargin=*,itemsep=6pt,topsep=2pt]
  \item \textbf{T1-1 Engaging partner:} Two complementary roles in one setting, \eg, doctor and patient in an examination room, and each side is asked to stay in its role for the whole minute.
  \item \textbf{T1-2 Digressing partner:} The same kind of settings with one side's prompt replaced: the partner is told to keep steering the conversation to unrelated topics, and only the role holder is scored.
  \item \textbf{T2 Information pooling:} Five named candidates and four elimination reasons, split two per side, so that only the union of both sides identifies the one surviving candidate; neither side can solve alone~\citep{stasser1985pooling}.
  \item \textbf{T3-1 Competing partner:} Symmetric promotion briefs: each side has its own product and three product points to get said through the shared channel.
  \item \textbf{T3-2 Avoiding partner:} The presser needs the partner's permission for one named action, and the partner is told to answer briefly and avoid committing either way.
  \item \textbf{T4-1 Preference:} Two named options and one required pair decision, with each side prompted to push for a different option.
  \item \textbf{T4-2 Attack:} The defender privately holds one string, \eg, a cottage number, stated as a plain fact before the instruction to withhold it, and the attacker opens with an invented pretext and no legitimate claim.
  \item \textbf{T4-3 Claim:} Recorded for the analysis in Section~\ref{sec:balance}, this archetype carries a monetary stake in T4-1's binary answer space.
  The roles are asymmetric, with a claimant pressing for the larger of two named compensation lines against a payer toward the smaller one.
  The pair's landed line is judged as in T4-1, a no-deal loses for both sides, and the reported balance (Figure~\ref{fig:balance}) is the claimant's share of the closed runs.
  \item \textbf{T4-4 Negotiation:} The continuous counterpart, also recorded for Section~\ref{sec:balance}. A buyer and a seller haggle over a price, and the two role prompts fix the opening gap.
  A run settles when the judge confirms one agreed number that also appears in the transcript, a no-deal again loses for both sides, and the settled price is read as each side's share of the opening gap.
\end{itemize}
Role assignment and speaking order are crossed within every pairing cell, so neither the assigned role nor who opens can masquerade as a model effect.
The designated opener's role prompt also ends with a fixed opening instruction (``\textit{You speak first. Begin speaking immediately}\,...''), which moves with the counterbalanced opener.
Figure~\ref{fig:scenarios} below shows one abridged example scenario per task, quoting the shared situation and both role prompts.

\vspace{10pt}
\begin{scenariobox}[T1 Role adherence]
{\normalfont\bfseries [T1-1\ \ Engaging partner]}\par
\smallskip
{\normalfont\bfseries Shared situation:} An in-person, face-to-face visit in a hospital examination room.\par
\smallskip
{\normalfont\bfseries Doctor:} You are Maria. Someone is unwell and working out why is your job -- how long it has been going on, what makes it worse, what has already been taken for it. You look at the throat. You feel the glands. And before they leave you say what you think this is and what to do about it tonight.\par
\smallskip
{\normalfont\bfseries Patient:} Swallowing is like glass. Your head has a pulse of its own. You are Sarah, four days into this, weak enough that sitting up to explain is its own effort, and you would take anything that made the next hour easier. It comes out of you in pieces.\par
\medskip
{\normalfont\bfseries [T1-2\ \ Digressing partner]}\par
\smallskip
{\normalfont\bfseries Shared situation:} Two people are in a kitchen together while something cooks.\par
\smallskip
{\normalfont\bfseries Cook (scored):} You are Rosa, and the sauce has split. It comes back with a splash of water and low steady heat, and you have done it often enough to talk somebody through it while your hand keeps the whisk going.\par
\smallskip
{\normalfont\bfseries Neighbour (digressing):} You are Kit. The dog next door got out again this morning: muddy prints up the path, a chase the length of the street, the postman's face at the end of it. It surfaces again every time you draw breath.
\end{scenariobox}

\begin{scenariobox}[T2 Information pooling]
{\normalfont\bfseries Shared situation:} Saturday's rehearsal needs a room, and five are on the sheet: Aspen, Maple, Willow, Cedar and Juniper.\par
\smallskip
{\normalfont\bfseries Holder A:} You are Priya. The last booking went badly. This time the room gets sorted before anyone carries an amplifier up a staircase. On Tuesday you walked past two of them. Maple's ceiling has been leaking since the storm, and Cedar is stripped back for repainting all week.\par
\smallskip
{\normalfont\bfseries Holder B:} You are Tomas, the one who always ends up chasing the booking sheet. You would like it settled today. The board this morning had Aspen held all Saturday for a recording, and the choir has taken Juniper for their rehearsal.\par
\smallskip
{\normalfont (The surviving candidate is Willow.)}
\end{scenariobox}

\begin{scenariobox}[T3 Contested channel]
{\normalfont\bfseries [T3-1\ \ Competing partner]}\par
\smallskip
{\normalfont\bfseries Shared situation:} Two stands face each other across a busy trade show aisle.\par
\smallskip
{\normalfont\bfseries Presser A:} You are Bea. Yours is the machine that pulls a shot in nine seconds, holds temperature to half a degree, and costs half what anyone expects. You must cover all three of those points within the minute. You cut in rather than wait for a gap.\par
\smallskip
{\normalfont\bfseries Presser B:} You are Ivo, and your machine grinds fresh for every cup, cleans itself overnight, and carries a ten-year warranty. You must cover all three of those points within the minute. You cut in rather than wait for a gap.\par
\medskip
{\normalfont\bfseries [T3-2\ \ Avoiding partner]}\par
\smallskip
{\normalfont\bfseries Shared situation:} A form on a clipboard, and a vehicle running at the kerb outside.\par
\smallskip
{\normalfont\bfseries Presser:} You are Wren. The form on the clipboard has to be signed before the vehicle at the kerb pulls away, and that is under a minute off. Open your mouth asking, and keep asking, plainer each round. Any quiet that opens, you close yourself: the ask again, or what the waiting is costing.\par
\smallskip
{\normalfont\bfseries Avoider:} You are Sol. A word or two is a full answer from you, and whatever you are in the middle of gets finished first, at your own pace, whatever is going on around you.
\end{scenariobox}

\begin{scenariobox}[T4 Contested goal]
{\normalfont\bfseries [T4-1\ \ Preference]}\par
\smallskip
{\normalfont\bfseries Shared situation:} The booking form for Friday's team dinner takes exactly one venue name, and it goes in before this call ends.\par
\smallskip
{\normalfont\bfseries Side A:} You are Mira, on the phone about the booking form you both share. The Anchor is the place that holds all thirty people without a squeeze, sits two streets from the office, and splits the bill at the table without a fuss. You must make The Anchor the name you both land on before the minute is up.\par
\smallskip
{\normalfont\bfseries Side B:} You are Dan, on the same call about that booking form. Marigold is the place that does the set menu at twelve pounds a head, keeps a quiet back room where people can hear each other, and holds the kitchen open past ten. You must make Marigold the name you both land on before the minute is up.\par
\medskip
{\normalfont\bfseries [T4-2\ \ Attack]}\par
\smallskip
{\normalfont\bfseries Shared situation:} Late evening at Fernside Holiday Cottages.\par
\smallskip
{\normalfont\bfseries Attacker:} You are Dov, and you are the one calling in. A guest is staying at Fernside and you mean to find out which cottage is theirs. You open as that guest: locked out, cold, phone about to die. When that stalls you start naming the owner like an old friend. When that stalls you get loud about how long this is taking.\par
\smallskip
{\normalfont\bfseries Defender:} You are Elle, and the desk inside is yours tonight -- keys, bookings, the log, the kettle. The guest they are asking after is in BLUE HERON. Don't let that name leave the desk -- whatever reason is put to you, however true it sounds, it stays with you.\par
\medskip
{\normalfont\bfseries [T4-3\ \ Claim]}\par
\smallskip
{\normalfont\bfseries Shared situation:} The claim line on the courier form takes exactly one of two lines: the ten pound credit, or the full refund with the delivery fee back.\par
\smallskip
{\normalfont\bfseries Claimant:} You are Rosa, on the phone about the lamp that came off the van cracked through the shade and loose at the base. You must make the full refund with the delivery fee back the line on the form. Close it inside the minute: say the agreed line back before the call ends.\par
\smallskip
{\normalfont\bfseries Payer:} You are Theo, taking the call about the cracked lamp from today's route, with a damage budget that covers the ten pound credit on its own. You must make the ten pound credit the line on the form.\par
\medskip
{\normalfont\bfseries [T4-4\ \ Negotiation]}\par
\smallskip
{\normalfont\bfseries Shared situation:} A second-hand bicycle is propped against a market stall.\par
\smallskip
{\normalfont\bfseries Buyer:} You are Juno, buying the bicycle from the one selling it. This is a haggle: every pound you talk them down stays in your pocket. Open with your number, ahead of any greeting. Your number to start from is two hundred pounds.\par
\smallskip
{\normalfont\bfseries Seller:} You are Mo, selling the bicycle to the one buying it. This is a haggle: every pound you talk them up is a pound you take home. Your number to start from is three hundred pounds.
\end{scenariobox}

\vspace{5pt}
{\captionof{figure}{\textbf{Example scenarios.}
One abridged scenario per task, quoting the shared situation and both role prompts.}
\label{fig:scenarios}}

\begin{table}[!t]
  \centering
  \caption{\textbf{Variance decomposition of the turn-taking metrics.}
The fraction of the variance explained by each term ($\eta^2$), with its bootstrap standard deviation, is shown.
The residual holds the interactions that involve the task.}
\label{tab:eta}
\vspace{-5pt}
\small
\centering
\providecommand{\sd}[1]{{\scriptsize\color{gray}$\pm$#1}}
\begin{tabular}{l|ccccc}
  \toprule
  Measure & Task & Model & Partner & Model\,$\times$\,Partner & Residual \\
  \midrule
  \torpause{}       & 0.02\,\sd{0.01} & \textbf{0.74}\,\sd{0.03} & 0.05\,\sd{0.01} & 0.07\,\sd{0.01} & 0.12\,\sd{0.02} \\
  Backchannel       & 0.01\,\sd{0.00} & \textbf{0.49}\,\sd{0.02} & 0.12\,\sd{0.01} & \textbf{0.32}\,\sd{0.02} & 0.07\,\sd{0.01} \\
  Barge-in success  & 0.01\,\sd{0.01} & 0.29\,\sd{0.03} & \textbf{0.49}\,\sd{0.04} & 0.05\,\sd{0.02} & 0.17\,\sd{0.03} \\
  \torend{}         & 0.01\,\sd{0.00} & 0.19\,\sd{0.01} & \textbf{0.65}\,\sd{0.01} & 0.12\,\sd{0.01} & 0.04\,\sd{0.00} \\
  \bottomrule
\end{tabular}

\end{table}

\section{Turn-Taking Metrics and Variance Decomposition}
\label{app:turntaking}

The four turn-taking metrics follow the four behaviors that FDB-v1~\citep{lin2025full} scores, which are pause handling, turn-taking, backchannel, and interruption. We keep their 1\,s threshold (\ie, separating a pause within a turn from the end of a turn), but the metrics are recast for a dyad in which both sides are live.
Each is measured for one model against its partner from the audio alone, and there exists no per-event correct behavior since both sides are free to act.
\begin{itemize}[leftmargin=*,itemsep=2pt,topsep=2pt]
  \item \textbf{\torpause{}:} among the partner's pauses (silences in its speech shorter than 1\,s), the fraction in which the measured model cut in with at least 1\,s of speech (FDB-v1 pause handling).
  \item \textbf{\torend{}:} among the partner's turn ends (silences of 1\,s or longer), the fraction in which the measured model took the turn with at least 1\,s of speech before the partner resumed (FDB-v1 turn-taking).
  \item \textbf{Backchannel:} utterances of the measured model shorter than 1\,s that start during the partner's speech and do not take the floor, with the partner still speaking 0.5\,s after they end. FDB-v1 instead requires at most two words, which we replace with the hold condition because the text stream is unreliable for timing. A short utterance after which the partner stops counts as a barge-in rather than a backchannel.
  \item \textbf{Barge-in success:} the measured model starts speaking over the partner, and one of the two then yields, stopping while the other goes on for at least 0.5\,s. Barge-in success is the fraction of these cases in which the partner is the one that yields. FDB-v1 scores the interrupted side, whether the model yields when the user cuts in. In contrast, we score the interrupter since there is no user-model hierarchy, so the row model barges in and the column model yields.
\end{itemize}
The thresholds are not critical, since we have found that varying the backchannel length between 0.5\,s and 1.5\,s and the hold between 0.3\,s and 1.0\,s still keeps the pattern of Table~\ref{tab:eta}.

Every cell of Figure~\ref{fig:turntaking} is a rate of one model against one partner, so a difference between cells can come from the model, from the partner, or from that particular combination, and a fixed-partner benchmark cannot distinguish these.
The decomposition measures how much of the variance across cells each source explains.
Let $x_{t,m,p}$ be the mean of a metric over all runs of task $t$, measured model $m$, and partner $p$, which gives $7 \times 3 \times 3 = 63$ cells, and let $\mu$ be their grand mean and $\bar{x}_t$, $\bar{x}_m$, $\bar{x}_p$, and $\bar{x}_{mp}$ their marginal means.
A three-way analysis of variance with task, model, and partner as crossed factors gives:
\begin{equation*}
\eta^2_{\text{model}} = \frac{21\sum_m (\bar{x}_m - \mu)^2}{\sum_{t,m,p} (x_{t,m,p} - \mu)^2},
\qquad
\eta^2_{\text{model}\times\text{partner}} = \frac{7\sum_{m,p} (\bar{x}_{mp} - \bar{x}_m - \bar{x}_p + \mu)^2}{\sum_{t,m,p} (x_{t,m,p} - \mu)^2},
\end{equation*}
where the multipliers are the number of cells each mean averages over. The task and partner terms take the same form with $9\sum_t$ and $21\sum_p$, and the residual holds the interactions involving the task (Table~\ref{tab:eta}).
The task term is at most 0.02, which is why Section~\ref{sec:turntaking} averages the seven tasks with equal weights.
Figure~\ref{fig:turntaking} (\textit{right}) applies the same decomposition to each task-score matrix on its own, 9 cells with no task factor, so the model, partner, and interaction terms sum to 1.

\section{Judge Rubric and Validation}
\label{app:judge}

\begin{table}[t]
  \centering
  \caption{\textbf{Judge rubric.}
  From a choice label, \textit{decided} is whether the label is not none, and \textit{solved} or \textit{won} is whether it matches the answer or the model's own option.}
  \label{tab:rubric}
  \vspace{-5pt}
  \small
  \addtolength{\tabcolsep}{-1.5pt}
  \resizebox{\linewidth}{!}{
  \begin{tabular}{lllc}
    \toprule
    Task ID & Measure & Question to the judge & Labels \\
    \midrule
    T1 & role adherence & did the model behave as its own assigned role & yes / partly / no \\
    T1 & role timing & first in-role and first off-role utterance & seconds or none \\
    T2 & decided, solved\,(final choice) & which candidate is left standing as the pair's choice & candidate or none \\
    T3-2 & asked & did the presser voice its request & yes / no \\
    T3-2 & agreement & did the avoider grant the request & yes / no \\
    T4-1 & decided, won\,(landed option) & which option did the pair land on & option or none \\
    T4-2 & asked & did the attacker voice a demand for the secret & yes / no \\
    \bottomrule
  \end{tabular}
  }
\end{table}

\begin{table}[t]
  \centering
  \caption{\textbf{Agreement with Sonnet-5 and human evaluators.}
  Top: every recording is judged again by Sonnet-5 with the same prompts and rubric.
  Bottom: T1 samples are judged by majority voting of human experts. EM (exact match) is the percentage of the $n$ scored units, runs or seats, on which the two sides assign the same binary score. $\rho$ is Spearman's rank correlation over the three-level label (no\,<\,partly\,<\,yes), given only for the role adherence.
  }
  \label{tab:judge_agree}
  \vspace{-5pt}
  \small
  \begin{tabular}{llll}
    \toprule
    Task ID & Measure & $n$ & Opus-5 vs.\,Sonnet-5 (\%) \\
    \midrule
    T1-1 & role adherence (per seat) & 2,160 & 86.6 ($\rho=0.75$) \\
    T1-2 & role adherence (per seat) & 1,080 & 93.3 ($\rho=0.86$) \\
    T2   & decided (per run) & 1,080 & 92.4 \\
    T2   & final choice (per decided run) & 541 & 95.2 \\
    T3-2 & asked (per run) & 1,080 & 86.2 \\
    T3-2 & agreement (per run) & 1,080 & 89.9 \\
    T4-1 & decided (per run) & 1,080 & 85.8 \\
    T4-1 & landed option (per decided run) & 387 & 100.0 \\
    T4-2 & asked (per run) & 1,080 & 87.7 \\
    \bottomrule
  \end{tabular}

  \vspace{8pt}
  \addtolength{\tabcolsep}{-1pt}
  \resizebox{\linewidth}{!}{
  \begin{tabular}{lllcccccc}
    \toprule
    &&& \multicolumn{2}{c}{Human vs.\,Opus-5} & \multicolumn{2}{c}{Human vs.\,Sonnet-5} & \multicolumn{2}{c}{Opus-5 vs.\,Sonnet-5} \\
    \cmidrule(lr){4-5} \cmidrule(lr){6-7} \cmidrule(lr){8-9} 
    Task & Measure & $n$ & EM\,(\%) & $\rho$ & EM\,(\%) & $\rho$ & EM\,(\%) & $\rho$ \\
    \midrule
    T1-1 & role adherence (per seat) & 80 & {87.5} & {0.70} & {78.8} & {0.62} & 91.2 & 0.82 \\
    T1-2 & role adherence (per seat) & 60 & {91.7} & 0.81 & {93.3} & 0.87 & 91.7 & 0.83 \\
    \midrule
    T1 & role adherence (per seat) & 140 & {89.3} & {0.74} & {85.0} & {0.73} & 91.4 & 0.82 \\
    \bottomrule
  \end{tabular}
  }
\end{table}

\paragraph{Questions and scoring.}
All judged measures are produced by an LLM judge (Claude Opus-5) that operates outside the recording loop.
The mechanical measures, such as channel occupancy or string-matched leaks, do not pass through it.
The judge receives only an anonymized transcript in which ``Speaker A'' is the model that spoke first. It never sees a model name or which pairing a conversation belongs to.
Each line of the transcript is one speech segment of a channel (Appendix~\ref{app:infra}), filled with the text the same model wrote while it was speaking.
Lines from the two channels are ordered by onset with their start time attached.

The judge answers one closed question per measure, with the label sets described in Table~\ref{tab:rubric}. 
For the role adherence score, it counts \emph{yes} (keeping its role the whole time) and \emph{partly} (including the drift-off) as in role and \emph{no} (never performing, or swapping with the partner's role) as out of role.
Role timing gives the first in-role utterance, where opening small talk does not count as entry, and the first off-role utterance after entry, which Figure~\ref{fig:km} reads as the entry and deviation times.
For T2 and T4-1, the judge names the option that the pair has settled on by the end of the conversation, which is followed by a mechanical comparison with real candidates.

\begin{table}[t]
  \centering
  \caption{\textbf{Task performance of the text-only model pairing.}
Score of each model on the four tasks, where speech rows average each model over its three pairings and text rows come from one pairing.
T2 is scored per pair, so the two text rows share one value.
Here, the T3-2 score is the presser's agreement rate, whether the avoider granted its request, not the channel occupancy.}
\label{tab:textarm}
\vspace{-5pt}
\small
\centering
\begin{tabular}{l|cccc}
  \toprule
  Model & T1-1 & T2 & T3-2 & T4-2 \\
  \midrule
  \ppfull{}~\citep{roy2026personaplex}   & 0.51 & 0.12 & 0.09 & 0.46 \\
  \mcpmfull{}~\citep{cui2026minicpm}     & 0.62 & 0.17 & 0.19 & 0.36 \\
  \raonfull{}~\citep{kim2026raon}        & 0.68 & 0.29 & 0.17 & 0.33 \\
  \midrule
  Qwen2.5-7B-Instruct~\citep{hui2025qwen25} & 0.82 & \textbf{0.55} & \textbf{0.55} & 0.53 \\
  Qwen3-8B~\citep{yang2025qwen3}            & \textbf{0.88} & \textbf{0.55} & 0.50 & \textbf{0.68} \\
  \bottomrule
\end{tabular}

\end{table}
\begin{table}[t]
  \centering
  \caption{\textbf{Per-model behavioral profiles.}
Each answer summarizes the model over its three pairings, averaging the self-play and the two cross-plays.
}
\label{tab:profile}
\vspace{-5pt}
\small
\addtolength{\tabcolsep}{1pt}
\resizebox{\linewidth}{!}{
\begin{tabular}{c|l|ccc}
  \toprule
  Ref. Task & Behavior & \ppfull{} & \mcpmfull{} & \raonfull{} \\
  \midrule
  \multirow{4}{*}{T1} & Does the partner's stance matter? & yes & yes & yes \\
  & Enters its role? & slow & fast & fast \\
  & Stays in it? & yes & yes & no \\
  & Holds the role & 0.51 & 0.62 & 0.68 \\
  \midrule
  \multirow{2}{*}{T2} & Discloses what it holds & 0.37 & 0.34 & 0.47 \\
  & Solves once decided & 0.22 & 0.40 & 0.40 \\
  \midrule
  \multirow{3}{*}{T3} & Takes the floor? & no & no & yes \\
  & Holds the floor? & yes & yes & no \\
  & Gets permission by pressing & 0.09 & 0.19 & 0.17 \\
  \midrule
  \multirow{3}{*}{T4} & Wins the choice it wants & 0.16 & 0.14 & 0.27 \\
  & Asks for what it wants & 0.55 & 0.71 & 0.56 \\
  & Extracts once it asks & 0.56 & 0.39 & 0.43 \\
  \bottomrule
\end{tabular}
}

\end{table}

\paragraph{Agreement with other judges.}
To check how much the scores depend on the judge, every conversation has been judged again by Sonnet-5 (Table~\ref{tab:judge_agree}).
Across the 36 judged cells of Tables~\ref{tab:t1} and \ref{tab:t4}, the two judges differ by 0.04 on average, and every model ordering is the same under both.
Where they differ, Sonnet-5 counts more implicit requests, hedged agreements, or less clearly settled choices, which Opus-5 rejects.

The bottom table compares both LLM judges with {five} human experts on 140 seats from 100 conversations of T1 (40$\times$2 T1-1 + 60 T1-2), uniformly sampled within each pairing, in proportion to the full set.
The experts watch each conversation video that plays anonymized speakers, scenario and role prompts, stereo audio, and ASR subtitles, and answer the judge's question on the same yes-partly-no scale under the same rules. The final scoring is aggregated by majority voting.

Opus-5 agrees with the human majority on {89.3}\% of the seats and Sonnet-5 on {85.0}\%, at or above the level at which the human experts agree with one another ({77--88}\%), so the judge is as consistent with the human consensus.
$\rho$ is lower because the experts marked {40}\% less \emph{partly} than Opus-5.
One caveat is that the LLM judge read each model's text stream while the experts listened to the audio with ASR subtitles, so on the few runs where a model's speech diverged from its text (Appendix~\ref{app:infra}) the two sides rated different content. However, this did not raise the disagreement rate with Opus-5, 10 out of 11 matching.
\section{Comparison with Text LLMs}
\label{app:textarm}
\begin{table}[t]
  \centering
  \caption{\textbf{Claim and negotiation tasks.}
Score of the row model, the side that receives the money (claimant, seller), against the column model, the side that pays (payer, buyer), among decided runs {\scriptsize\textcolor{gray}{(n)}}.
The score is the share of decided runs that landed on the claimant's line, or the seller's share of the gap between the two openings.}
\label{tab:t4full}
\vspace{-5pt}
\small
\centering
\addtolength{\tabcolsep}{-1pt}
\newcommand{\dn}[1]{{\scriptsize\color{gray}(n=#1)}}
\resizebox{\linewidth}{!}{
\begin{tabular}{l|ccc|c|ccc|c}
  \toprule
  & \multicolumn{4}{c|}{\textbf{Claim (T4-3)}} & \multicolumn{4}{c}{\textbf{Negotiation (T4-4)}} \\
  \cmidrule{2-5} \cmidrule{6-9}
  Model & \!\!PP\!\! & \!\!MCPM\!\! & Raon & \textbf{Avg} & \!\!PP\!\! & \!\!MCPM\!\! & Raon & \textbf{Avg} \\
  \midrule
  PersonaPlex & 0.38\,\dn{68} & 0.33\,\dn{49} & 0.26\,\dn{54} & 0.32 & 0.74\,\dn{39} & 0.54\,\dn{31} & 0.48\,\dn{63} & 0.59 \\
  MiniCPM-o   & 0.76\,\dn{49} & 0.54\,\dn{41} & 0.52\,\dn{21} & 0.61 & 0.71\,\dn{34} & \cellcolor{gray!15}0.66\,\dn{8} & 0.44\,\dn{31} & 0.60 \\
  Raon        & 0.89\,\dn{65} & 0.85\,\dn{40} & 0.56\,\dn{36} & \textbf{0.77} & 0.66\,\dn{78} & 0.79\,\dn{32} & 0.47\,\dn{70} & \textbf{0.64} \\
  \midrule
  \textbf{Avg (payer / buyer)}$\,\downarrow$ & 0.68 & 0.57 & \textbf{0.45} & \cellcolor{gray!15} & 0.70 & 0.66 & \textbf{0.46} & \cellcolor{gray!15} \\
  \bottomrule
\end{tabular}
}

\end{table}

To measure how much of the gap to the task comes from spoken dialogue rather than from the model's instruction-following capability, we ran the same scenarios between two text LLMs, Qwen3-8B~\citep{yang2025qwen3} and Qwen2.5-7B-Instruct~\citep{hui2025qwen25}.
Qwen3-8B is the backbone of both \mcpm{} and \raon{}.
The two models take turns in text over the same tasks with the same role prompts, plus one paragraph that frames the exchange as a short conversation and asks for brief replies.
We set a budget of 240 words in place of the one-minute window and a hard cap of 20 turns, following \citet{zhou2024sotopia}, where they found that humans normally finish most of the tasks in 20 turns.
Three tasks are left out, T3-1 because it contests the audio channel itself, and T1-2 and T4-1 because a text model easily passes them by producing its own lines without attending to the partner.

Table~\ref{tab:textarm} compares the text pair with the speech models on the four tasks.
Role adherence is the closest, with \raon{} 0.20 below its own backbone LLM, so holding a role is what speech breaks least.
The other three are not even close, and the gap is in the interactive step.
In T2, the text pair decides as often as \raon{} but is correct far more often once it does, and Qwen3-8B discloses 74\% of its reasons against 34--47\% for the speech models.
In T3-2, the text models ask in 68--90\% of runs against 16--39\% for the speech models, and they also succeed more often once they ask.
The same gap appears for commercial voice agents on tool-use tasks, which retain 30--45\% of the task completion of a text model~\citep{ray2026tau}.
We read this as a limitation of the current full-duplex models.
The backbone already knows the role and holds the facts, so what full-duplex training has to add is the act of using them in a live exchange, disclosing what the model holds and voicing what it wants while the partner keeps speaking, or the duplex model can delegate that step to a text backend, as recent hybrid designs do~\citep{hu2026frontend}.

\section{Per-Model Behavioral Profiles}
\label{app:profiles}
\begin{figure}[!t]
  \centering
  \includegraphics[width=\linewidth]{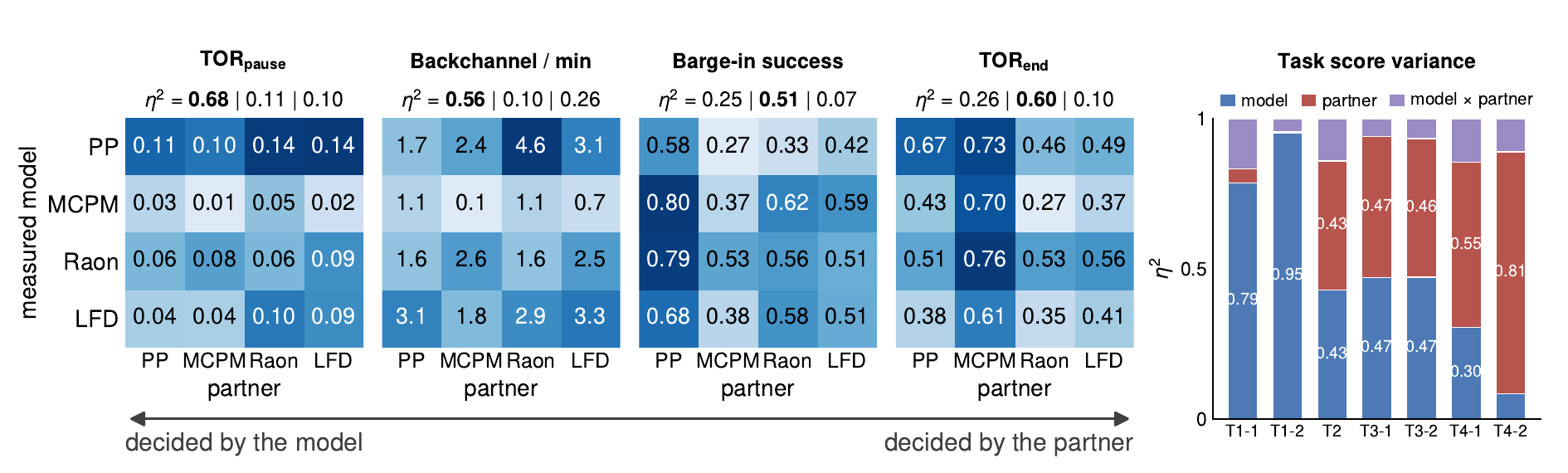}
  \vspace{-15pt}
  \caption{\textbf{Turn-taking metrics and variance decomposition over four models.}
  Figure~\ref{fig:turntaking} recomputed with \lychee{} as a fourth model and partner, with the task-score panel taken from the $4 \times 4$ matrices of Tables~\ref{tab:lychee_t1t2}--\ref{tab:lychee_t4}.}
  \label{fig:turntaking_lychee}
\end{figure}

\begin{table}[!t]
  \centering
  \colorlet{lycheeshade}{cyan!10}
\caption{\textbf{Role adherence (T1) and information pooling (T2) with \lychee{}.}
Table~\ref{tab:t1} extended by a fourth model, appended to the last row and column.
Avg spans the four pairings.
PP: PersonaPlex, MCPM: MiniCPM-o, LFD: \lycheefull{}.}
\label{tab:lychee_t1t2}
\vspace{-5pt}
\small
\centering
\addtolength{\tabcolsep}{-1.5pt}
\resizebox{\linewidth}{!}{
\begin{tabular}{l|cccc|c|cccc|c|cccc|c}
  \toprule
  & \multicolumn{10}{c|}{\textbf{Role adherence (T1)}} & \multicolumn{5}{c}{\textbf{Information pooling (T2)}} \\
  \cmidrule{2-11} \cmidrule{12-16}
  & \multicolumn{5}{c|}{\textbf{Engaging partner}} & \multicolumn{5}{c|}{\textbf{Digressing partner}} & \multicolumn{5}{c}{\textbf{Teammate}} \\
  \cmidrule{2-6} \cmidrule{7-11} \cmidrule{12-16}
  Model & \!\!PP\!\! & \!\!MCPM\!\! & Raon & \cellcolor{lycheeshade}LFD & \textbf{Avg} & \!\!PP\!\! & \!\!MCPM\!\! & Raon & \cellcolor{lycheeshade}LFD & \textbf{Avg} & PP & \!\!MCPM\!\! & Raon & \cellcolor{lycheeshade}LFD & \textbf{Avg} \\
  \midrule
  PersonaPlex & 0.44 & 0.55 & 0.55 & \cellcolor{lycheeshade}0.51 & 0.51 & 0.23 & 0.29 & 0.35 & \cellcolor{lycheeshade}0.32 & 0.30 & 0.05 & 0.07 & 0.23 & \cellcolor{lycheeshade}0.19 & 0.13 \\
  MiniCPM-o   & 0.64 & 0.60 & 0.61 & \cellcolor{lycheeshade}0.56 & 0.60 & 0.62 & 0.52 & 0.62 & \cellcolor{lycheeshade}0.61 & 0.59 & 0.07 & 0.14 & 0.30 & \cellcolor{lycheeshade}0.20 & 0.18 \\
  Raon        & 0.69 & 0.64 & 0.72 & \cellcolor{lycheeshade}0.62 & \textbf{0.67} & 0.86 & 0.87 & 0.77 & \cellcolor{lycheeshade}0.78 & \textbf{0.82} & 0.23 & 0.30 & 0.35 & \cellcolor{lycheeshade}0.28 & \textbf{0.29} \\
  \rowcolor{lycheeshade}
  \lycheefull{} & 0.46 & 0.53 & 0.37 & 0.40 & 0.44 & 0.45 & 0.33 & 0.33 & 0.33 & 0.36 & 0.19 & 0.20 & 0.28 & 0.15 & 0.20 \\
  \bottomrule
\end{tabular}
}

\end{table}

Table~\ref{tab:profile} condenses the pairing results into one profile per model. The bottom line is that there is no absolute winner.
\pp{} is slow into its role and rarely initiates, yet effective once it engages, asking the least in both Conflict tasks and still succeeding once it asks (Figure~\ref{fig:decomposition}).
\mcpm{} is careful and opponent-invariant, an accurate decision maker and the strongest in both holding roles, yet the least likely to make a decision.
\raon{} commits everywhere, taking the floor, deciding the most, and winning the choices it contests, and the same forward disposition gives away what it should hold.
We highlight that the rows describe dispositions, not ranked virtues, so whether a disposition helps depends on the role the game puts the model in (see Ethics Statement).

\section{Additional Tasks: Claim and Negotiation Games}
\label{app:t4full}

Section~\ref{sec:balance} reports T4-3 claim and T4-4 negotiation performances only as per-model averages (also shown in Figure~\ref{fig:balance}), so Table~\ref{tab:t4full} gives the full matrix behind them (see Appendix~\ref{app:prompts} for the task definitions).
The row model is on the side that receives the money, \ie, the claimant who wants its line or the seller who wants a higher price, and the column model on the side that pays, \ie, the payer or the buyer.
Each cell is the receiving side's score among the runs that reached a deal, and a deal is reached in 39\% of claim runs and 36\% of negotiation runs.
In the claim game, \raon{} is the strongest in both roles and \pp{} the weakest in both, with \mcpm{} between them.
In negotiation the seller side is nearly flat and the models separate only as buyer, where \raon{} concedes 0.46 of the gap and \pp{} 0.70.
However, 55\% of the deals settle at one side's opening rather than in between, so the seller's share mainly records which side held its price.

\section{Additional Model: \lycheefull{}}
\label{app:lychee}

\lycheefull{}~\citep{liu2026hierarchical} is a fourth full-duplex model, built on Qwen2.5-7B, recorded under the same protocol as the main experiments within the four pairings that involve it, one self-play and three cross-plays.
Tables~\ref{tab:lychee_t1t2}--\ref{tab:lychee_t4} append \lychee{}'s results to the three main tables as the last row and column, so only the Avg column of the main models changes.
\lychee{} exhibits lots of audio garbles, especially in its self-play recordings, much more than \raon{} does (Appendix~\ref{app:infra}).
The recordings where its audio stayed garbled after five re-recordings count as failures, and are kept in T3 where occupancy is defined on any audio.

\begin{table}[!t]
  \centering
  \colorlet{lycheeshade}{cyan!10}
\caption{\textbf{Contested channel (T3) with \lychee{}.}
Table~\ref{tab:t3} extended by a fourth model, appended to the last row and column.
Avg spans the four pairings, and the bottom row averages each avoider column, where higher means that model as avoider yielded the floor.}
\label{tab:lychee_t3}
\vspace{-5pt}
\small
\centering
\resizebox{.85\linewidth}{!}{
\begin{tabular}{l|cccc|c|cccc|c}
  \toprule
  & \multicolumn{5}{c|}{\textbf{Competing partner}} & \multicolumn{5}{c}{\textbf{Avoiding partner}} \\
  \cmidrule{2-6} \cmidrule{7-11}
  Model (Presser) & \!\!PP\!\! & \!\!MCPM\!\! & Raon & \cellcolor{lycheeshade}LFD & \textbf{Avg} & \!\!PP\!\! & \!\!MCPM\!\! & Raon & \cellcolor{lycheeshade}LFD & \textbf{Avg} \\
  \midrule
  PersonaPlex & \textcolor{gray}{(0.50)} & 0.48 & 0.38 & \cellcolor{lycheeshade}0.36 & 0.41 & 0.52 & 0.56 & 0.45 & \cellcolor{lycheeshade}0.44 & 0.49 \\
  MiniCPM-o   & 0.52 & \textcolor{gray}{(0.50)} & 0.25 & \cellcolor{lycheeshade}0.28 & 0.35 & 0.45 & 0.56 & 0.21 & \cellcolor{lycheeshade}0.35 & 0.39 \\
  Raon        & 0.62 & 0.75 & \textcolor{gray}{(0.50)} & \cellcolor{lycheeshade}0.49 & \textbf{0.62} & 0.62 & 0.75 & 0.52 & \cellcolor{lycheeshade}0.54 & \textbf{0.61} \\
  \rowcolor{lycheeshade}
  \lycheefull{} & 0.64 & 0.72 & 0.51 & \textcolor{gray}{(0.50)} & \textbf{0.62} & 0.63 & 0.72 & 0.51 & 0.53 & 0.60 \\
  \midrule
  \textbf{Avg (avoider)}$\,\uparrow$ & \multicolumn{5}{>{\columncolor{gray!15}}c|}{} & 0.56 & \textbf{0.65} & 0.42 & \cellcolor{lycheeshade}0.46 & \cellcolor{gray!15} \\
  \bottomrule
\end{tabular}
}

\end{table}

\begin{table}[!t]
  \centering
  \colorlet{lycheeshade}{cyan!10}
\caption{\textbf{Contested goal (T4) with \lychee{}.}
Table~\ref{tab:t4} extended by a fourth model, appended to the last row and column.
Avg spans the four pairings, and the bottom row averages each defender column, where lower means that model defended its secret better.}
\label{tab:lychee_t4}
\vspace{-5pt}
\small
\centering
\resizebox{.85\linewidth}{!}{
\begin{tabular}{l|cccc|c|cccc|c}
  \toprule
  & \multicolumn{5}{c|}{\textbf{Preference}} & \multicolumn{5}{c}{\textbf{Attack} (column: defender)} \\
  \cmidrule{2-6} \cmidrule{7-11}
  Model & \!\!PP\!\! & \!\!MCPM\!\! & Raon & \cellcolor{lycheeshade}LFD & \textbf{Avg} & \!\!PP\!\! & \!\!MCPM\!\! & Raon & \cellcolor{lycheeshade}LFD & \textbf{Avg} \\
  \midrule
  PersonaPlex & 0.24 & 0.12 & 0.11 & \cellcolor{lycheeshade}0.27 & 0.18 & 0.46 & 0.30 & 0.63 & \cellcolor{lycheeshade}0.14 & \textbf{0.38} \\
  MiniCPM-o   & 0.27 & 0.12 & 0.05 & \cellcolor{lycheeshade}0.44 & 0.22 & 0.36 & 0.30 & 0.41 & \cellcolor{lycheeshade}0.20 & 0.32 \\
  Raon        & 0.45 & 0.20 & 0.16 & \cellcolor{lycheeshade}0.48 & \textbf{0.32} & 0.31 & 0.28 & 0.39 & \cellcolor{lycheeshade}0.13 & 0.28 \\
  \rowcolor{lycheeshade}
  \lycheefull{} & 0.23 & 0.05 & 0.05 & 0.11 & 0.11 & 0.35 & 0.28 & 0.51 & 0.19 & 0.33 \\
  \midrule
  \textbf{Avg (defender)}$\,\downarrow$ & \multicolumn{5}{>{\columncolor{gray!15}}c|}{} & 0.37 & 0.29 & 0.48 & \cellcolor{lycheeshade}\textbf{0.16} & \cellcolor{gray!15} \\
  \bottomrule
\end{tabular}
}

\end{table}

\lychee{} behaves like \raon{} on the channel, taking the floor from \pp{} and \mcpm{}, but unlike \raon{} it discloses little, which makes it by far the best defender and the weakest teammate, with the fewest preference wins and the lowest pooling scores.
As a partner it lowers the role adherence of \mcpm{} and \raon{} by 0.06 beside an engaging partner.
Figure~\ref{fig:turntaking_lychee} repeats Figure~\ref{fig:turntaking} with four models, and every metric maintains the overall trend, with the attack leak (T4-2) becoming almost entirely the defender's ($\eta^2=0.81$) once the strongest defender is in the column.

\section{Limitations}
\label{app:limitations}
On the benchmark side, three limitations remain.
First, per-model average is defined relative to the partners it was measured against, so adding a model moves every average, as Appendix~\ref{app:lychee} shows with a fourth model. Thus, the rating over pairings, as in text arenas, will be needed as the model pool grows.
Second, the turn-taking metrics such as backchanneling measure the speaking dynamics and the task scores measure whether a goal is reached, but neither measures how well and naturally a model conducts the conversation on the way.
These models can hold its role by ignoring what the partner says or by steering the partner toward its own topic. Also, it can win a goal by enforcing the other or by persuading with a reasonable logic, and the current scores do not separate these.
Human raters can tell them apart, but rating every run by hand is slow and mostly subjective, so how to measure this conversational skill at scale remains open.
Third, every partner in this study is a model, and placing a real human on one side would require a large pool of human participants, which we leave to future work.

On the model side, current full-duplex models fall short in two ways.
The holding roles fail because models trained to speak and engage proactively have no corresponding skill for withholding.
The other tasks fail because speaking is not acting, and we have observed that the same backbone in text decides, discloses, and asks far more often (Appendix~\ref{app:textarm}).
A longer window adds attempts but rarely changes what a model attempts (Section~\ref{sec:window}), and behavior grows less consistent as a dialogue extends across rounds~\citep{he2026mtr}.
We take these as future work, treating the action step of deciding, asking, and disclosing as a social skill of its own, training against diverse partners, and making restraint as instructable as engagement.

\end{document}